\documentclass[12pt,onecolumn]{IEEEtran}
\usepackage[latin1]{inputenc}
\usepackage{graphicx,amssymb,amsmath}
\usepackage{multicol}
\usepackage[noadjust]{cite}
\usepackage[justification=centering]{caption}
\usepackage{setspace}
\usepackage{subfigure}
\usepackage{graphicx}
\usepackage{float}
\usepackage {url}
\usepackage{stfloats}
\usepackage{amsthm,pifont}
\usepackage{flushend}
\usepackage{cases,subeqnarray}
\usepackage{bm,multirow,bigstrut}
\usepackage{amsmath, amsthm, amssymb}
\usepackage{textcomp}
\usepackage{latexsym,bm}
\usepackage{booktabs}
\usepackage{xcolor}
\usepackage{mathtools}
\usepackage{dsfont}
\usepackage{extarrows}
\usepackage{epsfig}
\usepackage{epsfig}
\usepackage{epstopdf}
\usepackage[noend]{algpseudocode}
\usepackage{algorithmicx,algorithm}
\theoremstyle{plain}
\newtheorem{thm}{Theorem}

\theoremstyle{plain}
\newtheorem{rem}{Remark}
\newtheorem{cor}{Corollary}
\IEEEoverridecommandlockouts

\begin{document}


\title{Impact of Channel Aging on Cell-Free Massive MIMO Over Spatially Correlated Channels}

\author{Jiakang Zheng, Jiayi~Zhang,~\IEEEmembership{Senior Member,~IEEE}, Emil~Bj\"{o}rnson,~\IEEEmembership{Senior Member,~IEEE}, and Bo Ai,~\IEEEmembership{Senior Member,~IEEE}
\thanks{J. Zheng and J. Zhang are with the School of Electronics and Information Engineering, Beijing Jiaotong University, Beijing 100044, P. R. China. (e-mail: \{20111047, jiayizhang\}@bjtu.edu.cn).}
\thanks{E. Bj\"{o}rnson is with the Department of Electrical Engineering, Link\"{o}ping University, Link\"{o}ping, Sweden, and the Department of Computer Science, KTH Royal Institute of Technology, Kista, Sweden. (e-mail: emilbjo@kth.se).}
\thanks{B. Ai is with the State Key Laboratory of Rail Traffic Control and Safety, Beijing Jiaotong University, Beijing 100044, China. (e-mail: boai@bjtu.edu.cn)}
}

\maketitle
\vspace{-1cm}
\begin{abstract}
In this paper, we investigate the impact of channel aging on the performance of cell-free (CF) massive multiple-input multiple-output (MIMO) systems with both spatial correlation and pilot contamination. We derive novel closed-form uplink and downlink spectral efficiency (SE) expressions that take imperfect channel estimation into account. More specifically, we consider large-scale fading decoding and matched-filter receiver cooperation in the uplink. The uplink performance of a small-cell (SC) system is derived for comparison. The CF massive MIMO system achieves higher 95\%-likely uplink SE than the SC system. In the downlink, the coherent transmission has four times higher 95\%-likely per-user SE than the non-coherent transmission. Statistical channel cooperation power control (SCCPC) is used to mitigate the inter-user interference. SCCPC performs better than full power transmission, but the benefits are gradually weakened as the channel aging becomes stronger. Furthermore, strong spatial correlation reduces the SE but degrades the effect of channel aging. Increasing the number of antennas can improve the SE while decreasing the energy efficiency. Finally, we use the maximum normalized Doppler shift to design the SE-improved length of the resource block. Simulation results are presented to validate the accuracy of our expressions and prove that the CF massive MIMO system is more robust to channel aging than the SC system.
\end{abstract}
\IEEEpeerreviewmaketitle
\section{Introduction}
Cell-free (CF) massive multiple-input multiple-output (MIMO) has been recently proposed as a future technology for providing a more uniform spectral efficiency (SE) to the user equipments (UEs) in wireless networks \cite{zhang2020prospective,Ngo2017Cell}. CF massive MIMO systems consist of many geographically distributed access points (APs) connected to a central processing unit (CPU) for coherently serving the UEs by spatial multiplexing on the same time-frequency resource \cite{8000355,9186090,9184916}. The characteristic feature of CF massive MIMO, compared with traditional cellular systems, is the operating regime with no cell boundaries and many more APs than UEs \cite{bjornson2019making}. In conventional small-cell (SC) systems, the APs only serve UEs within their own cell, which may lead to high inter-cell interference.
In addition, compared with SC systems in mobile scenarios, CF massive MIMO systems can handle interference more effectively and the frequent handover problem can be eliminated \cite{zhang2020prospective}.
Results in \cite{Ngo2017Cell} and \cite{bjornson2019making} show that CF massive MIMO systems outperform SC systems in terms of 95\%-likely per-user uplink and downlink SE.
Following these seminal works, many important and fundamental aspects of CF massive MIMO have been studied in recent years. For example, the authors in \cite{nayebi2017precoding} observed that CF massive MIMO systems with the large-scale fading decoding (LSFD) receiver achieve two-fold gains over the matched filter (MF) receiver in terms of 95\%-likely per-user SE.
The key difference between LSFD and MF receivers is that the former requires more statistical parameters to be available at the CPU \cite{bjornson2019making,zheng2020efficient}.
In addition, it was shown in \cite{qiu2020downlink} that coherent transmission (same data from all APs) performs much better than non-coherent transmission (different data from all APs) in the downlink of CF massive MIMO. The main advantage of non-coherent transmission is that it does not require strict phase synchronization, which reduces the complexity of the system \cite{interdonato2019ubiquitous}.
Moreover, the authors in \cite{9079911} use the minimum mean-squared error-based successive interference cancellation scheme to detect the desired symbols for improving the performance of CF massive MIMO systems.
In addition, non-orthogonal pilot sequences are often used by the UEs in CF massive MIMO systems on account of the short training phases and large number of UEs \cite{9178782}. However, the pilot contamination effect caused by non-orthogonal pilot sequences reduces the performance of CF massive MIMO systems significantly, owing to the interfered pilot signal transmitted from other UEs assigned the same pilot \cite{mai2018pilot}.

Moreover, spatial channel correlation is an important property of multiuser MIMO \cite{fozooni2019hybrid}. The propagation channels and antenna arrays create spatial channel correlation, which has a non-negligible impact on the performance of CF massive MIMO \cite{polegre2020channel}. Spatially correlated Rayleigh fading channels were considered in CF massive MIMO in \cite{qiu2020downlink,bjornson2019making}. Rician fading channels with spatial correlation were investigated in \cite{polegre2020channel,jin2020spectral}.
It is not clear whether a UE at a given location will achieve higher SE with uncorrelated fading or with spatial correlation. Simulation results in \cite{bjornson2017massive} show that this happens with 17\%-35\% probability, which changes depending on the setup.

Higher SE and less consumed power are two of the key performance indicators for future networks \cite{9145564,9103348}. CF massive MIMO can potentially achieve a higher SE by deploying a large number of APs, but might increase the circuit power consumption \cite{bjornson2015optimal}. Moreover, more fronthaul links are required in CF massive MIMO, which also increase the total power consumption \cite{9124715}. Hence, it is hard to predict if CF massive MIMO will increase or decrease the energy efficiency (EE).
The authors in \cite{ngo2017total} analyzed the effects of fronthaul power consumption, the number of APs, and the number of antennas per AP on the EE of CF massive MIMO. The uplink EE of the CF Massive MIMO with optimal uniform quantization is investigated in \cite{bashar2019energy}. APs were temporarily turned off in \cite{van2020joint} to reduce the hardware dissipation, which can minimize the total downlink power consumption at the APs to achieve an EE improvement.

Most of the current works on CF massive MIMO consider a block-fading model, i.e., the channel realization in a coherence block is approximated as constant.
However, practical channels are continuously evolving due to UE mobility, leading to the so-called channel aging effect where the channel is different but correlated between samples in a transmission block \cite{truong2013effects}.
The impact of channel aging has been characterized in co-located massive MIMO systems. For instance, the authors in \cite{yuan2020machine} utilized machine learning to predict the channels in massive MIMO systems under channel aging. The performance of massive MIMO systems under channel aging was also investigated in \cite{chopra2018performance,papazafeiropoulos2016impact}.
In addition, the authors in \cite{8716688} utilized a low-complexity prediction to improve the channel state information accuracy in massive MIMO-orthogonal frequency division multiplexing (OFDM) systems with channel aging.
To the best of our knowledge, this is the first time that the channel aging effect is analyzed for CF massive MIMO systems.

Motivated by the above observations, we investigate the effect of channel aging on the uplink and downlink performance of CF massive MIMO systems with spatial correlation and pilot contamination. In the uplink, we consider both LSFD and MF receivers. The performance of the corresponding SC system is analyzed for comparison. In addition, we compare the coherent and non-coherent transmission modes in the downlink. Then, a useful statistical channel cooperation power control (SCCPC) scheme is applied to further improve the system performance. The total EE and the SE-improved length of the resource block of the considered system are investigated. The specific contributions of the work are listed as follows:
\begin{itemize}
  \item We first derive closed-form expressions for the uplink and downlink SE of the CF massive MIMO system under channel aging. Our results show that, in both static and mobile scenarios, the CF massive MIMO system performs better than the SC system in the uplink. In the downlink, coherent transmission performs better than non-coherent transmission.
  \item Compared with full power transmission, we find that the benefits of SCCPC are gradually weakened as the channel aging effect grows stronger. The performance loss due to channel aging can be compensated by employing more antennas, more pilots, and reducing the spatial correlation.
  \item We investigate the total EE of the CF massive MIMO system taking into account a realistic power consumption model. It is found that increasing channel aging significantly reduces the EE and leads to more APs are preferred for the optimal operating point of EE. We also propose a method to design the SE-improved length of resource block for mitigating the effect of channel aging.
\end{itemize}

Note that the conference version of this paper \cite{zheng2020cell} considered the uplink CF massive MIMO with channel aging under single-antenna APs, while this paper considers multi-antenna APs in both uplink and downlink.
The rest of the paper is organized as follows. In Section \ref{se:model}, we describe the system model incorporating the combined effects of channel estimation error, spatial correlation, pilot contamination and channel aging. Next, Section \ref{se:performance} presents the achievable uplink SE with channel aging for both CF and SC systems, as well as SCCPC is used to further improve the system performance. In Section \ref{se:downlink}, we derive the achievable downlink SE with channel aging for coherent and non-coherent transmissions and use SCCPC to further improve the system performance. Then, Section \ref{se:TEE} investigate the total EE of considered systems taking into account a realistic power consumption model. We provide numerical results and discussions in Section \ref{se:NR}. Finally, Section \ref{se:CON} gives a brief summary and provides suggestions for future work.

\textbf{Notation:} Column vectors and matrices are represented by boldface lowercase letters $\mathbf{x}$ and boldface uppercase letters $\mathbf{X}$, respectively.
The $n\times n$ identity matrix is ${{\mathbf{I}}_n}$.
We use superscripts $x^\mathrm{*}$, $\mathbf{x}^\mathrm{T}$ and $\mathbf{x}^\mathrm{H}$ to \mbox{represent} conjugate, transpose and conjugate transpose, respectively.
We use ${\mathrm{diag}}\left( {{x_1}, \ldots ,{x_n}} \right)$ for a block-diagonal matrix with the \mbox{variables} ${{x_1}, \ldots ,{x_n}}$ on the diagonal.
The absolute value, the expectation operator, the trace operator and the definitions are denoted by $\left|  \cdot  \right|$, $\mathbb{E}\left\{  \cdot  \right\}$, ${\text{tr}}\left(  \cdot  \right)$ and $\triangleq$, respectively.
Finally, $\mathbf{x} \sim \mathcal{C}\mathcal{N}\left( {\mathbf{0},\mathbf{R}} \right)$ represents a circularly symmetric complex Gaussian distribution with covariance matrix $\mathbf{R}$.

\section{System Model}\label{se:model}

\begin{figure}[t]
\centering
\includegraphics[scale=0.55]{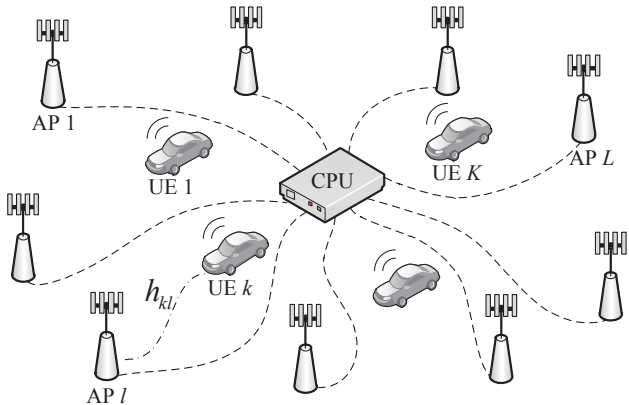}
\caption{CF massive MIMO systems with mobile UEs.} \vspace{-4mm}
\label{system_model}
\end{figure}

\begin{figure}
\centering
\includegraphics[scale=0.6]{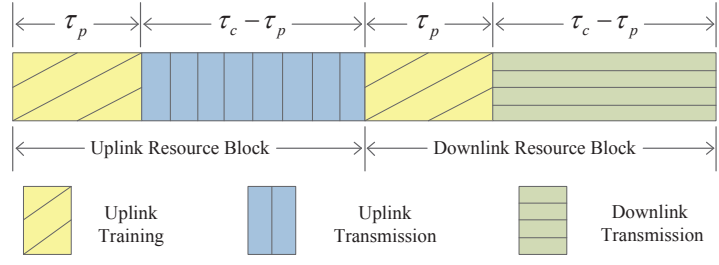}
\caption{Uplink and downlink resource blocks.}
\label{block}
\end{figure}

We consider a CF massive MIMO system consisting of $L$ APs and $K$ mobile single-antenna UEs as illustrated in Fig.~\ref{system_model}.
Each AP is equipped with $N$ antennas. The APs are connected to a CPU via fronthaul links. We assume that all $L$ APs simultaneously serve all $K$ UEs on the same time-frequency resource. The UEs are assumed to move at different speeds, which affect the channel variations.
The communication is divided into resource blocks consisting of $\tau_c$ time instants (channel uses).\footnote{Note that a resource block is not the same as a coherence block in the block fading model often considered in the literature \cite{bjornson2017massive}. We consider a more realistic model where the channels are continuously evolving in a resource block and correlated between channel uses but not identical.}
As the time-division duplex (TDD) protocol in Fig.~\ref{block} indicates, there are two different types of blocks: one for uplink data and one for downlink data.\footnote{On account of the channel aging effect, the estimated channel information will become outdated over time, which makes the system performance get worse over time. Therefore, we transmit uplink and downlink data in different resource blocks, so both can make use of the parts of the resource blocks where the channel information is most accurate. Another benefit is that this leads to less frequent switching between uplink and downlink, which reduces the number of guard intervals.} We assume that the uplink training phase occupies $\tau_p$ time instants, while the uplink data transmission occupies ${\left( {{\tau _c} - {\tau _p}} \right)}$ time instants in the uplink resource block. For the downlink resource block, the uplink training phase occupies $\tau_p$ time instants and the downlink data transmission occupies ${\left( {{\tau _c} - {\tau _p}} \right)}$ time instants.
The fraction of the resource blocks that are used for uplink and downlink data can be dynamically changed and will not be specified in this paper since each block is operated independently from the other.
We are considering a flat-fading system, which could either be a narrowband single-carrier system or one narrowband subcarrier in a multi-carrier system.
At the $n$th time instant in a given block, the Rayleigh fading channel between AP $l$ and UE $k$ is modelled as
\begin{align}
{{\mathbf{h}}_{kl}}\left[ n \right] \sim {\mathcal{C}\mathcal{N}}\left( {{\mathbf{0}},{{\mathbf{R}}_{kl}}} \right),\ n =0,1, \ldots ,{\tau _c},
\end{align}
where ${{\mathbf{R}}_{kl}} \in {\mathbb{C}^{N \times N}}$ is the spatial correlation matrix and $\beta _{kl} \triangleq {\text{tr}}\left( {{{\mathbf{R}}_{kl}}} \right)/N$ is the large-scale fading coefficient. Note that ${\mathbf{h}_{kl}}\left[ n \right]$ is independent for different pairs of UE index $k=1, \ldots ,K$ and AP index $l=1, \ldots ,L$. However, ${\mathbf{h}_{kl}}\left[ 0 \right], \ldots, {\mathbf{h}_{kl}}\left[ \tau_c \right]$ are correlated.

\subsection{Channel Aging}
The relative movement between the UEs and APs lead to temporal variations in the propagation environment which affect the channel coefficient also within a resource block.
The channel realization ${\mathbf{h}_{kl}}\left[ {n } \right]$ can be modeled as a function of its initial state ${\mathbf{h}_{kl}}\left[ 0 \right]$ and an innovation component \cite{chopra2018performance}, such as
\begin{align}\label{channelaging}
{\mathbf{h}_{kl}}\left[ {n } \right] = {\rho _k}\left[ {n } \right]{\mathbf{h}_{kl}}\left[ 0 \right] + {\bar \rho_k }\left[ {n } \right]{\mathbf{g}_{kl}}\left[ {n} \right],
\end{align}
where ${\mathbf{g}_{kl}}\left[ {n } \right] \sim \mathcal{C}\mathcal{N}\left( {0,{{{\mathbf{R}}_{kl}}}} \right)$ represents the independent innovation component at the time instant $n$.
In addition, $\rho_k\left[ {n} \right]$ represents the temporal correlation coefficient of UE $k$ between the channel realizations at time $0$ and $n$, and ${{\bar \rho }_k}\left[ n \right] = \sqrt {1 - {\rho^2_k}\left[ n \right]} $.
As in \cite{chopra2018performance}, we consider ${\rho _k}\left[ n \right] = {J_0}\left( {2\pi {f_{D,k}}{T_s}n} \right)$, where ${J_0}\left(  \cdot  \right)$ is the zeroth-order Bessel function of the first kind \cite[Eq. (9.1.18)]{Abramowitz1964table}, $T_s$ denotes the sampling time, and $f_{D,k}=(v_k f_c)/c$ is the Doppler shift for a UE with velocity $v_k$, where $f_c$ and $c$ represent the carrier frequency and the speed of light, respectively.
Our focus is on the channel aging and combatting it using time correlation.
\begin{rem}
The model in \eqref{channelaging} is not a first-order autoregressive model, as considered in previous works such as \cite{truong2013effects}, but the correlation coefficients are selected to approximately match the Jakes' model which makes it more realistic \cite{chopra2018performance}.
In practice, the model \eqref{channelaging} is only accurate during a limited period of time since wide-sense stationarity can only be guaranteed for small-scale movements. To avoid this issue, we are not exploiting correlation between blocks, even if it might exist.
\end{rem}

\subsection{Uplink Channel Estimation}

We assume that $\tau_p$ mutually orthogonal time-multiplexed pilot sequences are utilized. This means that pilot sequence $t$ corresponds to sending a pilot signal only at time instant $t$ in the resource block. This pilot design is necessary to keep the orthogonality between the pilots in the presence of channel aging.
We consider a large network with $K>\tau_p$ so that different UEs will be assigned to the same time instant (same pilot).
The index of the time instant assigned to UE $k$ is denoted by ${t_k} \in \left\{ {1, \ldots ,{\tau _p}} \right\}$ and the other UEs that use the same time instant for pilot transmission as UE $k$ is denoted by ${\mathcal{P}_k} = \{ i : t_i = t_k \} \subset \left\{ {1, \ldots ,K} \right\}$.
The received signal between AP $l$ and UE $k$ at time instant $t_k$ is
\begin{align}\label{zl}
{\mathbf{z}_l}\left[ t_k \right] = \sum\limits_{i \in {\mathcal{P}_k}} {\sqrt {{p_i}} {\mathbf{h}_{il}}\left[ t_i \right]} + {\mathbf{w}_l}\left[ t_k \right],
\end{align}
where ${p_i} \geqslant 0$ is the pilot transmit power of UE $i$ and ${{\mathbf{w}}_l}\left[ t_k \right] \sim {\mathcal{CN}}\left( {\mathbf{0},{\sigma^2\mathbf{I}_N}} \right)$ is the receiver noise.
This received signal can be utilized to estimate (or predict) the channel realization at any time instant of the block, but the quality of the estimate will reduce with an increased time distance between the pilot transmission and the considered channel realization. Without loss of generality, we consider the estimates at the channels at time instant $\tau_p+1$, and then these estimates are used as the initial states to get estimates of the channels at all other time instants.
To simplify the notation, we define $\lambda  = {\tau _p} + 1$.
The effective channel at the $t_i$th time instant can be expressed in terms of the channel at the $\lambda$th time instant as
\begin{align} \label{eq:relation-different-instants}
{\mathbf{h}_{il}}\left[ t_i \right]={\rho _i}\left[ {\lambda -t_i} \right]{\mathbf{h}_{il}}\left[ {\lambda } \right]+{{\bar \rho }_i}\left[ {\lambda -t_i} \right]{\mathbf{f}_{il}}\left[ t_i \right],
\end{align}
where ${\mathbf{f}_{il}}\left[ {t_i } \right] \sim \mathcal{C}\mathcal{N}\left( {\mathbf{0},{\mathbf{R}_{il}}} \right)$ denotes the independent innovation component that relate ${\mathbf{h}_{il}}\!\left[ t_i \right]$ and ${\mathbf{h}_{il}}\!\left[ {\lambda} \right]$. Using \eqref{eq:relation-different-instants}, we rewrite \eqref{zl} as
\begin{align}\label{z_l}
{\mathbf{z}_l}\left[ t_k \right] &=\sqrt {{p_k}}{\rho _k}\left[ {\lambda  - t_k} \right]{\mathbf{h}_{kl}}\left[ {\lambda } \right] + \!\!\!\!\sum\limits_{{i \in {\mathcal{P}_k} / \{ k \} }} \!\!\!{\sqrt {{p_i}} {\rho _i}\left[ {\lambda  - t_i} \right]{\mathbf{h}_{il}}\left[ {\lambda } \right]}  \notag\\
&+ \sum\limits_{i \in {\mathcal{P}_k}} {\sqrt {{p_i}} {{\bar \rho }_i}\left[ {\lambda  - t_i} \right]{\mathbf{f}_{il}}\left[ t_i \right]}  + {\mathbf{w}_l}\left[ t_k \right].
\end{align}
Using standard minimum mean square error (MMSE) estimation \cite{bjornson2017massive}, each AP $l$ can compute the MMSE estimate ${{\mathbf{\hat h}}_{kl}}\left[ {\lambda } \right]$ of the channel coefficient ${\mathbf{h_{kl}}}\left[ {\lambda } \right]$ as
\begin{align}\label{hhat}
{{{\mathbf{\hat h}}}_{kl}}\left[ \lambda  \right] = {\rho _k}\left[ {\lambda  - {t_k}} \right]\sqrt {{p_k}} {{\mathbf{R}}_{kl}}{\mathbf{\Psi }}_{kl}{{\mathbf{z}}_l}\left[ {{t_k}} \right],
\end{align}
where
\begin{align}
{{\mathbf{\Psi }}_{kl}} = {\left( {\sum\limits_{i \in {\mathcal{P}_k}} {{p_i}{{\mathbf{R}}_{il}}}  + {\sigma ^2}{{\mathbf{I}}_N}} \right)^{ - 1}}.
\end{align}
The estimate ${{\mathbf{\hat h}}_{kl}}\left[ {\lambda } \right]$ and the estimation error ${{{\mathbf{\tilde h}}}_{kl}}\left[ {\lambda } \right]=\mathbf{h}_{kl}\left[ {\lambda } \right]-{{\mathbf{\hat h}}_{kl}}\left[ {\lambda } \right]$ are distributed as $\mathcal{C}\mathcal{N}\left( {\mathbf{0},{{\mathbf{Q}}_{kl}}} \right)$ and $\mathcal{C}\mathcal{N}\left( {\mathbf{0},{{\mathbf{R}}_{kl}} - {{\mathbf{Q}}_{kl}}} \right)$, respectively,
where
\begin{align} \label{eq:variance-estimate}
{{\mathbf{Q}}_{kl}} \triangleq \rho _k^2\left[ {\lambda - {t_k}} \right]{p_k}{{\mathbf{R}}_{kl}}{{\mathbf{\Psi }}_{kl}}{{\mathbf{R}}_{kl}}.
\end{align}
In addition, to simplify the notation, we define
\begin{align}\label{Q_hat}
{{{\mathbf{\bar Q}}}_{kil}} \triangleq {\rho _k}\left[ {\lambda  - {t_k}} \right]\sqrt {{p_k}} {\rho _i}\left[ {\lambda  - {t_i}} \right]\sqrt {{p_i}} {{\mathbf{R}}_{il}}{{\mathbf{\Psi }}_{kl}}{{\mathbf{R}}_{kl}}.
\end{align}
The channel estimate ${{\mathbf{\hat h}}_{kl}}\left[ {\lambda } \right]$ is degraded by the signals transmitted by the pilot-sharing UEs at the time instant $t_k$. This represents the pilot contamination effect \cite{Ngo2017Cell}.
When ${{\mathbf{R}}_{kl}} = {\beta _{kl}}{{\mathbf{I}}_N}$, the derived channel estimate ${{{\mathbf{\hat h}}}_{kl}}\left[ \lambda  \right]$ can be reduced to the uncorrelated Rayleigh fading. Thus, it can be simplified as
\begin{align}
{{{\mathbf{\hat h}}}_{kl}}\left[ \lambda  \right] = {\rho _k}\left[ {\lambda  - {t_k}} \right]\sqrt {{p_k}} {\beta _{kl}}{\Psi _{kl}}{{\mathbf{I}}_N}{{\mathbf{z}}_l}\left[ {{t_k}} \right],
\end{align}
where
\begin{align}
{\Psi _{kl}} = {\left( {\sum\limits_{i \in {\mathcal{P}_k}} {{p_i}{\beta _{il}}}  + {\sigma ^2}} \right)^{ - 1}}.
\end{align}
Furthermore, we have that ${{\mathbf{Q}}_{kl}} \triangleq {\gamma _{kl}}{{\mathbf{I}}_N}$, where
\begin{align}
{\gamma _{kl}} = \rho _k^2\left[ {\lambda  - {t_k}} \right]{p_k}\beta _{kl}^2{\Psi _{kl}}.
\end{align}
Note that, when the channel aging becomes zero, the considered resource block becomes the conventional block-fading model. We find that the above expressions reduce to the channel estimation results in \cite{Ngo2017Cell} and \cite{bjornson2019making}, which are not influenced by the different time instants.

\section{Uplink Data Transmission}\label{se:performance}

In this section, we investigate the uplink performance of CF massive MIMO systems with channel aging and spatial correlation within each uplink resource block. Meanwhile, we consider SC systems for comparison. Novel SE expressions are derived for both systems and SCCPC is used to further improve the system performance.
\subsection{CF Massive MIMO Systems}

Each AP makes a local estimate of the uplink data using its local channel estimates. These data estimates are then sent to the CPU for joint data detection \cite{bjornson2019making}. During the uplink data transmission, the received complex baseband signal $\mathbf{y}_l[n]$ at AP $l$ during the instants $\lambda  \leqslant n \leqslant {\tau _c}$  is given by
\begin{align}\label{y_l}
{\mathbf{y}_l}\left[ {n } \right] = \sqrt{p_\text{u}} \sum\limits_{i = 1}^K {{\mathbf{h}_{il}}\left[ {n } \right]} \sqrt{\eta_i} {s_i}\left[ {n } \right] + {\mathbf{w}_l}\left[ {n } \right],
\end{align}
where $s_{i}\left[ {n } \right] \sim \mathcal{CN}\left(0, 1\right)$ is the transmit signal from UE $i$, and $0 \leqslant {\eta _i} \leqslant 1$ is a power control coefficient. $\mathbf{w}_{l} \left[ {n } \right]$ is the receiver noise. Moreover, ${\mathbf{h}_{il}}\left[ n \right]$ can be expressed in terms of $\mathbf{h}_{il} \left[\lambda  \right]$ as
\begin{align}\label{h_il}
{\mathbf{h}_{il}}\left[ n \right] = {\rho _i}\left[ {n - \lambda } \right]{\mathbf{h}_{il}} \left[\lambda  \right] + {{\bar \rho }_i}\left[ {n - \lambda } \right]{\mathbf{u}_{il}}\left[ n \right],
\end{align}
where $\mathbf{h}_{il}\left[\lambda  \right]$ is the initial state in the data transmission phase. The MMSE estimate ${{\mathbf{\hat h}}_{il}}\left[\lambda  \right] \sim{\mathcal{CN}}\left( {\mathbf{0}, {\mathbf{Q}_{il}} } \right)$ of $\mathbf{h}_{il}\left[\lambda  \right]$ was derived in \eqref{hhat}. Moreover, ${\mathbf{u}_{il}}\left[ {n } \right] \sim \mathcal{C}\mathcal{N}\left( {\mathbf{0},{\mathbf{R}_{il}}} \right)$ denotes the independent innovation component relating ${\mathbf{h}_{il}}\left[ n \right]$ and ${\mathbf{h}_{il}} \left[\lambda  \right]$.

To detect the symbol transmitted from the $k$th UE, the $l$th AP multiplies the received signal ${\mathbf{y}_l}\left[ {n } \right]$ with the conjugate of its (locally obtained) channel estimate.
Then the obtained quantity ${{\overset{\lower0.5em\hbox{$\smash{\scriptscriptstyle\smile}$}}{s} }_{kl}}\left[ n \right] \triangleq {\mathbf{\hat h}}_{kl}^{\text{H}}\left[ \lambda  \right]{{\mathbf{y}}_l}\left[ n \right]$ is sent to the CPU via the fronthaul. The CPU uses the weights $a_{kl} [n]$ to obtain ${{\hat s}_k}\left[ {n} \right]$ as
\begin{align}
  {{\hat s}_k}\left[ n \right] &= \sum\limits_{l = 1}^L {a_{kl}^ * \left[ n \right]{{\overset{\lower0.5em\hbox{$\smash{\scriptscriptstyle\smile}$}}{s} }_{kl}}} \left[ n \right] = \underbrace {{\rho _k}\left[ {n - \lambda } \right] \sqrt{p_\text{u}\eta_k} \sum\limits_{l = 1}^L {a_{kl}^ * \left[ n \right]\mathbb{E}\left\{ {{\mathbf{\hat h}}_{kl}^{\text{H}}\left[ \lambda  \right]{{\mathbf{h}}_{kl}}\left[ \lambda  \right]} \right\}} {s_k}\left[ n \right]}_{{\text{D}}{{\text{S}}_{k,n}}} \notag\\
   &+ \underbrace {{\rho _k}\left[ {n - \lambda } \right] \sqrt{p_\text{u}\eta_k} \left( {\sum\limits_{l = 1}^L {a_{kl}^ * \left[ n \right]\left( {{\mathbf{\hat h}}_{kl}^{\text{H}}\left[ \lambda  \right]{{\mathbf{h}}_{kl}}\left[ \lambda  \right] - \mathbb{E}\left\{ {{\mathbf{\hat h}}_{kl}^{\text{H}}\left[ \lambda  \right]{{\mathbf{h}}_{kl}}\left[ \lambda  \right]} \right\}} \right)} } \right){s_k}\left[ n \right]}_{{\text{B}}{{\text{U}}_{k,n}}} \notag\\
   &+ \underbrace {{{\bar \rho }_k}\left[ {n - \lambda } \right] \sqrt{p_\text{u}\eta_k} \sum\limits_{l = 1}^L {a_{kl}^ * \left[ n \right]{\mathbf{\hat h}}_{kl}^{\text{H}}\left[ \lambda  \right]{{\mathbf{u}}_{kl}}\left[ n \right]} {s_k}\left[ n \right]}_{{\text{C}}{{\text{A}}_{k,n}}} \notag \\
   &+  \sum\limits_{i \ne k}^K {\underbrace {\sqrt{p_\text{u}}\sum\limits_{l = 1}^L {a_{kl}^ * \left[ n \right]{\mathbf{\hat h}}_{kl}^{\text{H}}\left[ \lambda  \right]{{\mathbf{h}}_{il}}\left[ n \right] \sqrt{\eta_i} {s_i}\left[ n \right]} }_{{\text{U}}{{\text{I}}_{ki,n}}}}  + \underbrace {\sum\limits_{l = 1}^L {a_{kl}^ * \left[ n \right]{\mathbf{\hat h}}_{kl}^{\text{H}}\left[ \lambda  \right]{{\mathbf{w}}_l}\left[ n \right]} }_{{\text{N}}{{\text{S}}_{k,n}}},
\end{align}
where $\text{DS}_{k,n}$ represents the desired signal, $\text{BU}_{k,n}$ represents the beamforming gain uncertainty, $\text{CA}_{k,n}$ represents the channel aging effect,  $\text{UI}_{ki,n}$ represents the interference caused by transmitted data from other UEs, and $\text{NS}_{k,n}$ represents the noise term, respectively.

\begin{rem}
The channel aging effect clearly degrades both the desired signal and beamforming gain uncertainty since $\rho_k^2{\left[ {n - \lambda } \right]}$ reduces as $n$ increases. Focusing on the channel aging quantities, the SINR can be written approximately as $\text{SINR}_k\left[ {n} \right]\!\approx\! 1/\left( {a \kappa \! +\! b } \right)$, where $\kappa  \!\triangleq\! 1/{\rho_k^2{\left[ {n \!-\! \lambda } \right]}}$ and $a$, $b$ are constants depending on the transmit power, channel information, and receiver noise. It is clear that the SINR significantly reduces as the channel aging effect becomes large.
\end{rem}
\begin{thm}\label{th1}
A lower bound on the capacity of UE $k$ is
\begin{align}\label{SE_CF}
{\mathrm{SE}}_k^{{\mathrm{CF}}} = \frac{1}{{{\tau _c}}}\sum\limits_{n =\lambda}^{\tau_c} {{{\log }_2}\left( {1 + {\mathrm{SINR}}_k^{{\mathrm{CF}}}\left[ n \right]} \right)},
\end{align}
with
\begin{align}\label{SINR_k}
{\mathrm{SINR}}_k^{{\mathrm{CF}}}\!\left[ n \right] \!=\! \frac{{\rho _k^2\left[ {n - \lambda } \right]{{{p_{{\mathrm{u}}}}{\eta _k}}}{{\left| {{\mathbf{a}}_k^{\mathrm{H}}\left[ n \right]{{\mathbf{b}}_k}} \right|}^2}}}{{{p_{{\mathrm{u}}}}\sum\limits_{i = 1}^K {{\eta_i}{\mathbf{a}}_k^{\mathrm{H}}\!\left[ n \right]{{\mathbf{\Gamma }}_{ki}}{{\mathbf{a}}_k}\!\left[ n \right]}  \!+\! {p_{{\mathrm{u}}}}\!\!\!\sum\limits_{i \in {\mathcal{P}_k}\setminus \{ k \}} \!\!\!{\rho _i^2\!\left[ {n \!-\! \lambda } \right]{\eta_i}{{\left| {{\mathbf{a}}_k^{\mathrm{H}}\!\left[ n \right]{{\mathbf{c}}_{ki}}} \right|}^2}}  \!+\! {\sigma _{\mathrm{u}}^2}{\mathbf{a}}_k^{\mathrm{H}}\!\left[ n \right]{{\mathbf{\Lambda }}_k}{{\mathbf{a}}_k}\!\left[ n \right]}},
\end{align}
where
\begin{align}
  {{\mathbf{a}}_k}\left[ n \right] &\triangleq {\left[ {{a_{k1}}\left[ n \right] \ldots {a_{kL}}\left[ n \right]} \right]^{\mathrm{T}}} \in {\mathbb{C}^L} ,\notag \\
  {{\mathbf{b}}_k} &\triangleq {\left[ {{\mathrm{tr}}\left( {{{\mathbf{Q}}_{k1}}} \right) \ldots {\mathrm{tr}}\left( {{{\mathbf{Q}}_{kL}}} \right)} \right]^{\mathrm{T}}} \in {\mathbb{C}^L} ,\notag \\
  {{\mathbf{\Gamma }}_{ki}} &\triangleq {\mathrm{diag}}\left( {{\mathrm{tr}}\left( {{{\mathbf{Q}}_{k1}}{{\mathbf{R}}_{i1}}} \right), \ldots ,{\mathrm{tr}}\left( {{{\mathbf{Q}}_{kL}}{{\mathbf{R}}_{iL}}} \right)} \right) \in {\mathbb{C}^{L \times L}} ,\notag \\
  {{\mathbf{c}}_{ki}} &\triangleq {\left[ { {{\mathrm{tr}}\left( {{{{\mathbf{\bar Q}}}_{ki1}}} \right)} \ldots {{\mathrm{tr}}\left( {{{{\mathbf{\bar Q}}}_{kiL}}} \right)} } \right]^{\mathrm{T}}} \in {\mathbb{C}^L}, \notag \\
  \label{trace}{{\mathbf{\Lambda }}_k} &\triangleq {\mathrm{diag}}\left( {{\mathrm{tr}}\left( {{{\mathbf{Q}}_{k1}}} \right), \ldots ,{\mathrm{tr}}\left( {{{\mathbf{Q}}_{kL}}} \right)} \right) \in {\mathbb{C}^{L \times L}} .
\end{align}
\end{thm}
\begin{IEEEproof}
Please refer to Appendix A.
\end{IEEEproof}

We will call the lower bound on the capacity in Theorem~\ref{th1} an achievable SE. It can be achieved by using a set of $\left(\tau_c-\tau_p\right)$ channel codes for AWGN channels, each spanning over the signal transmitted at the $n$th time instant in every resource block, for $n= \lambda , \ldots ,{\tau _c}$.
The weight vector ${{\mathbf{a}}_k} [n]$ can be different for each $n$ and can be optimized by the CPU to maximize the SE, utilizing the LSFD receiver cooperation approach from \cite{nayebi2017precoding,bjornson2019making}. In addition, while the channel aging effect vanishes, we have $\rho _i^2\left[ {n - \lambda } \right] = 1$ at each time instant. Then, the achievable SE in Theorem~\ref{th1} will degenerate into the results in \cite[Proposition 2]{bjornson2019making}.

\begin{cor}
The effective SINR of UE $k$ is maximized by
\begin{align}
{{\mathbf{a}}_k} [n]= {\left( {{p_{{\mathrm{u}}}}\sum\limits_{i = 1}^K {{\eta_i}{{\mathbf{\Gamma }}_{ki}}}+ {p_{{\mathrm{u}}}}\sum\limits_{i \in {\mathcal{P}_k}\setminus \{ k \}}{\rho _i^2{\left[ {n-\lambda } \right]}{\eta_i}{{\mathbf{c}}_{ki}}{\mathbf{c}}_{ki}^{\mathrm{H}}}  + {\sigma _{\mathrm{u}}^2}{{\mathbf{\Lambda }}_k}} \right)^{ - 1}}{{\mathbf{b}}_k}, \notag
\end{align}
which leads to the maximum SE as
\begin{align}
{\mathrm{SE}}_k^{{\mathrm{LSFD}}} = \frac{1}{{{\tau _c}}}\sum\limits_{n =\lambda}^{\tau_c} {{{\log }_2}\left( {1 + {\mathrm{SINR}}_k^{{\mathrm{LSFD}}}\left[ n \right]} \right)},
\end{align}
where ${\mathrm{SINR}}_k^{{\mathrm{LSFD}}} [n]$ is given by
\begin{align}
 { \rho _k^2\left[ {n - \lambda } \right]{p_{{\mathrm{u}}}}{\eta_k}{\mathbf{b}}_k^{\mathrm{H}}{{\left( {{p_{{\mathrm{u}}}}\sum\limits_{i = 1}^K {{\eta_i}{{\mathbf{\Gamma }}_{ki}}}  + {p_{{\mathrm{u}}}}\sum\limits_{i \in {\mathcal{P}_k}\setminus \{ k \}} {\rho _i^2\left[ {n - \lambda } \right]{\eta_i}{{\mathbf{c}}_{ki}}{\mathbf{c}}_{ki}^{\mathrm{H}}}  + {\sigma _{\mathrm{u}}^2}{{\mathbf{\Lambda }}_k}} \right)}^{ - 1}}{{\mathbf{b}}_k}}.
\end{align}
\end{cor}
If we want to reduce the complexity of LSFD, then the conventional MF receiver cooperation from \cite{Ngo2017Cell} is obtained by using equal weights ${{\mathbf{a}}_k} [n] = {\left[ {1/L \ldots 1/L} \right]^{\mathrm{T}}}$.

\begin{cor}\label{cor_un}
For the spatially uncorrelated Rayleigh fading, substituting ${\mathrm{tr}}\left( {{{\mathbf{Q}}_{kl}}{{\mathbf{R}}_{il}}} \right) = N{\gamma _{kl}}{\beta _{il}}$, ${\mathrm{tr}}\left( {{{\mathbf{Q}}_{kl}}} \right) = N{\gamma _{kl}}$ and ${\mathrm{tr}}\left( {{{{\mathbf{\bar Q}}}_{kil}}} \right) = N\sqrt {{\gamma _{kl}}{\gamma _{il}}} $ into \eqref{trace}, we can write ${\mathrm{SINR}}_k^{{\mathrm{CF}}}\left[ n \right]$ as
\begin{align}\label{N_un}
 \frac{{\rho _k^2\left[ {n - \lambda } \right]{p_{\mathrm{u}}}{\eta _k}{N}{{\left| {\sum\limits_{l = 1}^L {a_{kl}^ * \left[ n \right]{\gamma _{kl}}} } \right|}^2}}}{{{p_{\mathrm{u}}}\!\sum\limits_{i = 1}^K {{\eta _i}\!\sum\limits_{l = 1}^L {{{\left| {a_{kl}^ * \!\left[ n \right]} \right|}^2}} {\gamma _{kl}}{\beta _{il}}}  \!+\! {p_{\mathrm{u}}}{N}\!\!\!\!\sum\limits_{i \in {\mathcal{P}_k}\setminus \{ k \}}\!\!\!\! {\rho _i^2\!\left[ {n \!-\! \lambda } \right]{\eta _i}{{\left| {\sum\limits_{l = 1}^L {a_{kl}^ * \!\left[ n \right]\!\sqrt {{\gamma _{kl}}{\gamma _{il}}} } } \right|}^2}} \!\!\! +\! \sigma _{\mathrm{u}}^2\!\sum\limits_{l = 1}^L {{{\left| {a_{kl}^ * \!\left[ n \right]} \right|}^2}{\gamma _{kl}}} }}.
\end{align}
\end{cor}
Corollary \ref{cor_un} is a general expression for CF massive MIMO systems with spatially uncorrelated multiple-antennas APs.
Assuming $N=1$ and applying MF receiver cooperation with ${{a}_{kl}} [n] = 1/L$, formula \eqref{N_un} will reduce to \cite[Eq. (27)]{Ngo2017Cell} when there is no channel aging effect.
\begin{rem}\label{Re_SE_N}
It is noticed that, in Corollary \ref{cor_un}, the number of antennas per AP have two impacts on the SINR. Firstly, the desired signal power grows proportionally to $N$. Secondly, the pilot-contaminated interference terms also grows linearly with $N$. Focusing $N$, the SINR at one fixed time instant can be written approximately as ${\mathrm{SINR}}_k^{{\mathrm{CF}}} = 1/\left( {\dot{a} \dot{\kappa}  + \dot{b} } \right)$, where $\dot{\kappa}  \triangleq 1/N$ and $\dot{a}$, $\dot{b}$ are positive constants. $\dot{a}$ is mainly depended on the interference from other UEs and the noise, $\dot{b}$ is mainly decided by the pilot contamination. It is clear that the SINR is a monotonely increasing function of $N$.
\end{rem}

\subsection{Small-Cell Systems}

In the uplink of a SC system, each AP first estimates the channels based on signals sent from the UEs, as described earlier. The so-obtained channel estimate of UE $k$ is used to multiply the received signal for detecting the desired signal. The combined uplink signal at the $l$th AP is
\begin{align}\label{small cell}
  &{y_{kl}}\left[ n \right] = \sqrt{p_\text{u}} \sum\limits_{i = 1}^K {{\mathbf{\hat h}}_{kl}^{\text{H}}\left[ \lambda  \right]{{\mathbf{h}}_{il}}\left[ n \right]} \sqrt{\eta_i}{s_i}\left[ n \right] + {\mathbf{\hat h}}_{kl}^{\text{H}}\left[ \lambda  \right]{{\mathbf{w}}_l}\left[ n \right] \notag \\
   &= {\rho _k}\left[ {n - \lambda } \right]\sqrt{p_\text{u}}{\mathbf{\hat h}}_{kl}^{\text{H}}\left[ \lambda  \right]{{{\mathbf{\hat h}}}_{kl}}\left[ \lambda  \right]\sqrt{\eta_k}{s_k}\left[ n \right] + \underbrace {{\rho _k}\left[ {n - \lambda } \right]\sqrt{p_\text{u}}{\mathbf{\hat h}}_{kl}^{\text{H}}\left[ \lambda  \right]{{{\mathbf{\tilde h}}}_{kl}}\left[ \lambda  \right]\sqrt{\eta_k}{s_k}\left[ n \right]}_{{{\text{I}}_{kl,n,1}}} \notag \\
   &+ \underbrace {{{\bar \rho }_k}\left[ {n - \lambda } \right]\sqrt{p_\text{u}}{\mathbf{\hat h}}_{kl}^{\text{H}}\left[ \lambda  \right]{{\mathbf{f}}_{kl}}\left[ n \right]\sqrt{\eta_k}{s_k}\left[ n \right]}_{{{\text{I}}_{kl,n,2}}} + \underbrace {\sqrt{p_\text{u}}\sum\limits_{i \ne k}^K {{\mathbf{\hat h}}_{kl}^{\text{H}}\left[ \lambda  \right]{{\mathbf{h}}_{il}}\left[ n \right]} \sqrt{\eta_i}{s_i}\left[ n \right]}_{{{\text{I}}_{kl,n,3}}} + {\mathbf{\hat h}}_{kl}^{\text{H}}\left[ \lambda  \right]{{\mathbf{w}}_l}\left[ n \right],
\end{align}
where the first term denotes the desired received signal from UE $k$, ${\mathbf{w}_{{l}}}\left[ {n} \right] \sim \mathcal{C}\mathcal{N}\left( {\mathbf{0},{\sigma^2}} {{\mathbf{I}}_N}\right)$ is the receiver noise at AP $l$. The remaining terms ${{{\mathrm{I}}_{kln,1}}}$, ${{{\mathrm{I}}_{kln,2}}}$ and ${{{\mathrm{I}}_{kln,3}}}$ are uncorrelated and represent interference caused by channel estimation errors, the channel aging effect, and data transmitted from other UEs.
\begin{cor}\label{thm_SC}
The SC system is a special case of CF massive MIMO when the LSFD weights are selected so that each user is only served by the SE-maximizing AP \cite{bjornson2019making}.
If the maximum-ratio combining is used, then the capacity of UE $k$ is lower bounded by
\begin{align}
{\mathrm{SE}}_k^{{\mathrm{SC}}} = \mathop {\max }\limits_{l \in \left\{ {1, \ldots ,L} \right\}} \frac{1}{{{\tau _c} }}\sum\limits_{n=\lambda}^{\tau_c} {\mathbb{E}\left\{ {{{\log }_2}\left( {1 + {\mathrm{SINR}}_{kl}^{{\mathrm{SC}}}\left[ n \right]} \right)} \right\}},
\end{align}
where ${\mathrm{SINR}}_{kl}^{{\mathrm{SC}}}\left[ n \right] $ is given by
\begin{align}
\frac{{\rho _k^2\left[ {n - \lambda } \right]{p_{\mathrm{u}}}{\eta_k}{{\left| {{\mathbf{\hat h}}_{kl}^{\mathrm{H}}\left[ \lambda  \right]{{{\mathbf{\hat h}}}_{kl}}\left[ \lambda  \right]} \right|}^2}}}{{\sum\limits_{i \ne k}^K {\rho _i^2\left[ {n \!-\! \lambda } \right]{p_{\mathrm{u}}}{\eta_i}{{\left| {{\mathbf{\hat h}}_{kl}^{\mathrm{H}}\left[ \lambda  \right]{{{\mathbf{\hat h}}}_{il}}\left[ \lambda  \right]} \right|}^2} \!+\! {\mathbf{\hat h}}_{kl}^{\mathrm{H}}\left[ \lambda  \right]\left( {\sum\limits_{i = 1}^K {{p_{\mathrm{u}}}{\eta_i}\left( {{{\mathbf{R}}_{il}} \!-\! \rho _i^2\left[ {n \!-\! \lambda } \right]{{\mathbf{Q}}_{il}}} \right) \!+\! {\sigma ^2}{{\mathbf{I}}_N}} } \right){{{\mathbf{\hat h}}}_{kl}}\left[ \lambda  \right]} }}.
\end{align}
\end{cor}
\begin{IEEEproof}
It follows similar steps in \cite[Theorem 4.1]{bjornson2019making} for Cellular mMIMO.
\end{IEEEproof}
Corollary \ref{thm_SC} is a general expression for SC system with an arbitrary number of antennas per AP.
When considering that special case $N=1$, the following result is obtained instead.
\begin{cor}
When $N=1$, the achievable SE of UE $k$ can be expressed in closed form as
\begin{align}\label{w}
{\mathrm{SE}}_k^{{\mathrm{SC}}} \!=\! \mathop {\max }\limits_{l \in \left\{ {1, \ldots ,L} \right\}} \frac{1}{{{\tau _c}}}\sum\limits_{n =\lambda}^{\tau_c} {\frac{{{e^{\frac{1}{{{w_{kl}}\left[ n \right]\left( {1 + {A_{kl}}\left[ n \right]} \right)}}}}{E_1}\left( {\frac{1}{{{w_{kl}}\left[ n \right]\left( {1 + {A_{kl}}\left[ n \right]} \right)}}} \right) \!-\! {e^{\frac{1}{{{w_{kl}}\left[ n \right]{A_{kl}}\left[ n \right]}}}}{E_1}\left( {\frac{1}{{{w_{kl}}\left[ n \right]{A_{kl}}\left[ n \right]}}} \right)}}{{\ln \left( 2 \right)}}} ,
\end{align}
where
\begin{align}
{A_{kl}}\left[ n \right] &= \sum\limits_{i \in {\mathcal{P}_k} \setminus \{ k \} }^K {{{\left( {\frac{{{\rho _i}\left[ {n - \lambda } \right]{\rho _i}\left[ {\lambda  - {t_i}} \right]{p_i}{\beta _{il}}}}{{{\rho _k}\left[ {n -\lambda } \right]{\rho _k}\left[ {\lambda  - {t_k}} \right]{p_k}{\beta _{kl}}}}} \right)}^2}}, \\
{w_{kl}}\left[ n \right] &= \frac{{\rho _k^2\left[ {n - \lambda } \right]{p_\mathrm{u}}\eta_k{\gamma _{kl}}}}{{p_\mathrm{u}\sum\limits_{i = 1}^K {{\eta_i}{\beta _{il}}}  - p_\mathrm{u}\sum\limits_{i \in {\mathcal{P}_k}}^K {\rho _i^2\left[ {n - \lambda } \right]{\eta_i}{\gamma _{il}}}  + {\sigma^2}}},
\end{align}
and ${E_1}\left( x \right) = \int_1^\infty  {\frac{{{e^{ - xu}}}}{u}} du$ denotes the exponential integral. The UE is served by the AP whose index $l$ is maximizing the expression in \eqref{w}.
\end{cor}
\begin{IEEEproof}
Please refer to Appendix B.
\end{IEEEproof}

\subsection{Uplink Statistical Channel Cooperation Power Control}
The uplink transmit powers must be selected to mitigate near-far effects.
We will extend the SCCPC policy presented in \cite{nikbakht2019uplink} to the considered system. In the SC system, the power control coefficient of UE $k$ in the cell $l_k$ is selected as
\begin{align}\label{FPC_SC}
{\eta _k} = \frac{{\min \left\{ {{\beta _{k{l_k}}}} \right\}}}{{{\beta _{k{l_k}}}}},k = 1, \ldots ,K.
\end{align}
SCCPC in CF massive MIMO systems depends on all the large-scale fading coefficient that involve a given UE, reflecting the effective connection between the UE and all the APs. Therefore, the power control coefficient of UE $k$ is given by
\begin{align}\label{FPC_CF}
{\eta _k} = \frac{{\min \left\{ {{\beta _k}} \right\}}}{\beta _{k}},k = 1, \ldots ,K.
\end{align}
where
\begin{align}
{\beta _k} = \sum\nolimits_{l = 1}^L {{\beta _{kl}}}.
\end{align}

\section{Downlink Data Transmission}\label{se:downlink}

In this section, we investigate the downlink performance of CF massive MIMO systems with channel aging and spatial correlation within each downlink resource block. We also derive novel SE expressions for coherent and non-coherent transmissions and use SCCPC to further improve the system performance.
\subsection{Coherent Downlink Transmission}

Each downlink resource block contains $\left( {{\tau _c} - {\tau _p}} \right)$ time instants reserved for downlink data.
We assume that the downlink transmission of CF massive MIMO adopts coherent joint transmission. It means each AP transmits to each UE and sends the same data symbol as the other APs.
Using maximum ratio precoding, the transmitted signal from AP $l$ at time instant $n$ is
\begin{align}
{{\mathbf{x}}_l}\left[ n \right] = \sqrt{p_{\text{d}}}\sum\limits_{i = 1}^K {{{{\mathbf{\hat h}}}_{il}}\left[ \lambda  \right]{\sqrt {{\mu _{il}}} {q_i}\left[ n \right]}} ,
\end{align}
where ${q_{i}}\left[ n \right] \sim \mathcal{C}\mathcal{N}\left( {0,1} \right)$ is the symbol sent to UE $i$ which is same for all APs, ${p_{\text{d}}}$ denotes the downlink maximum transmission power for one AP, and
$0 \leqslant {\mu _{il}} \leqslant 1$ is power control coefficients chosen to satisfy the downlink power constraint as $\mathbb{E}\left\{ {{{\left| {{{\mathbf{x}}_l}\left[ n \right]} \right|}^2}} \right\} \leqslant {p_{\text{d}}}$.
With the help of \eqref{h_il}, the received signal at the $k$th UE at time instant $n$ is expressed as
\begin{align}\label{coherent}
  &r_k^{{\text{coh}}}\left[ n \right] = \sum\limits_{l = 1}^L {{\mathbf{h}}_{kl}^{\text{H}}\left[ n \right]{{\mathbf{x}}_l}} \left[ n \right] + {w_k}\left[ n \right] = \sqrt{{p_{\text{d}}}}\sum\limits_{i = 1}^K {\sum\limits_{l = 1}^L {{\mathbf{h}}_{kl}^{\text{H}}\left[ n \right]{{{\mathbf{\hat h}}}_{il}}\left[ \lambda  \right]\sqrt {{\mu _{il}}} {q_i}\left[ n \right]} }  + {w_k}\left[ n \right] \notag \\
   &= {\rho _k}\left[ {n - \lambda } \right]\sqrt{{p_{\text{d}}}}\sum\limits_{l = 1}^L {{\mathbf{h}}_{kl}^{\text{H}}\left[ \lambda  \right]{{{\mathbf{\hat h}}}_{kl}}\left[ \lambda  \right]\sqrt {{\mu _{kl}}} {q_k}\left[ n \right]}  + {{\bar \rho }_k}\left[ {n - \lambda } \right]\sqrt{{p_{\text{d}}}}\sum\limits_{l = 1}^L {{\mathbf{u}}_{kl}^{\text{H}}\left[ n \right]{{{\mathbf{\hat h}}}_{kl}}\left[ \lambda  \right]\sqrt {{\mu _{kl}}} {q_k}\left[ n \right]}  \notag\\
   &+ \sqrt{p_{\text{d}}}\sum\limits_{i \ne k}^K {\sum\limits_{l = 1}^L {{\mathbf{h}}_{kl}^{\text{H}}\left[ n \right]{{{\mathbf{\hat h}}}_{il}}\left[ \lambda  \right]\sqrt {{\mu _{il}}} {q_i}\left[ n \right]} }  + {w_k}\left[ n \right],
\end{align}
where ${w_k}\left[ n \right] \sim \mathcal{C}\mathcal{N}\left( {0,\sigma _{\text{d}}^2} \right)$ is the receiver noise at UE $k$.
\begin{thm}\label{thm_d}
Based on the signal in \eqref{coherent}, the DL capacity of UE $k$ is lower bounded as
\begin{align}
{\mathrm{SE}}_k^{{\mathrm{coh}}} = \frac{1}{{{\tau _c}}}\sum\limits_{n =\lambda}^{\tau_c} {{{\log }_2}\left( {1 + {\mathrm{SINR}}_k^{{\mathrm{coh}}}\left[ n \right]} \right)} ,
\end{align}
where
\begin{align}\label{sinr_coh}
{\mathrm{SINR}}_k^{{\mathrm{coh}}}\!\left[ n \right] \!=\! \frac{{\rho _k^2\left[ {n - \lambda } \right]{p_{{\mathrm{d}}}}{{\left| {\sum\limits_{l = 1}^L {\sqrt {{\mu _{kl}}} \mathbb{E}\left\{ {{\mathbf{h}}_{kl}^{\mathrm{H}}\left[ \lambda  \right]{{{\mathbf{\hat h}}}_{kl}}\left[ \lambda  \right]} \right\}} } \right|}^2}}}{{{p_{{\mathrm{d}}}}\!\sum\limits_{i = 1}^K {\!\mathbb{E}\!\left\{\! {{{\left| {\sum\limits_{l = 1}^L {\!\sqrt {{\mu _{il}}} {\mathbf{h}}_{kl}^{\mathrm{H}}\!\left[ n \right]{{{\mathbf{\hat h}}}_{il}}\!\left[ \lambda  \right]} } \right|}^2}} \!\right\} \!-\! \rho _k^2\!\left[ {n \!-\! \lambda } \right]{p_{{\mathrm{d}}}}{{\left| {\sum\limits_{l = 1}^L {\!\sqrt {{\mu _{kl}}} \mathbb{E}\!\left\{\! {{\mathbf{h}}_{kl}^{\mathrm{H}}\!\left[ \lambda  \right]{{{\mathbf{\hat h}}}_{kl}}\!\left[ \lambda  \right]} \!\right\}} } \right|}^2} \!+ } {{\sigma _{\mathrm{d}}^2}}}}.
\end{align}
Computing every term of \eqref{sinr_coh}, ${\mathrm{SINR}}_k^{{\mathrm{coh}}}\left[ n \right] $ can be further expressed in closed form as
\begin{align}\label{sinr_coh_closed}
{\mathrm{SINR}}_k^{{\mathrm{coh}}}\left[ n \right] \!=\! \frac{{\rho _k^2\left[ {n - \lambda } \right]{p_{{\mathrm{d}}}}{{\left| {\sum\limits_{l = 1}^L {\sqrt {{\mu _{kl}}} {\mathrm{tr}}\left( {{{\mathbf{Q}}_{kl}}} \right)} } \right|}^2}}}{{{p_{{\mathrm{d}}}}\!\sum\limits_{i = 1}^K {\!\sum\limits_{l = 1}^L {{\!\mu _{il}}} {\mathrm{tr}}\!\left( {{{\mathbf{Q}}_{il}}{{\mathbf{R}}_{kl}}} \right)}  \!+\! \rho _k^2\left[ {n \!-\! \lambda } \right]{p_{{\mathrm{d}}}}\!\!\!\sum\limits_{i \in {\mathcal{P}_k}\setminus \{ k \}} {{{\left| {\sum\limits_{l = 1}^L {\!\sqrt {{\mu _{il}}} {{\mathrm{tr}}\left( {{{{\mathbf{\bar Q}}}_{kil}}} \right)} } } \right|}^2}}  \!\!+\! {{\sigma _{\mathrm{d}}^2}}}}.
\end{align}
\end{thm}
\begin{IEEEproof}
It follows similar steps in Theorem \ref{th1}. According to similar methods of calculating \eqref{E_DS} and \eqref{E_UI} in Appendix A, we derive the closed-form expression for \eqref{sinr_coh} as \eqref{sinr_coh_closed}.
\end{IEEEproof}
Note that Theorem \ref{thm_d} is a general expression, which can degrade to \cite[Eq. (29)]{qiu2020downlink} as the channel aging disappears.
It is similar to Corollary \ref{cor_un}, substituting ${\mathrm{tr}}\left( {{{\mathbf{Q}}_{il}}{{\mathbf{R}}_{kl}}} \right) = N{\gamma _{il}}{\beta _{kl}}$, ${\mathrm{tr}}\left( {{{\mathbf{Q}}_{kl}}} \right) = N{\gamma _{kl}}$ and ${\mathrm{tr}}\left( {{{{\mathbf{\bar Q}}}_{kil}}} \right) = N\sqrt {{\gamma _{kl}}{\gamma _{il}}} $ into \eqref{sinr_coh_closed}, we can derive ${\mathrm{SINR}}_k^{{\mathrm{coh}}}\left[ n \right]$ with the spatially uncorrelated Rayleigh fading as
\begin{align}\label{sinr_coh_closed1}
{\mathrm{SINR}}_k^{{\mathrm{coh}}}\left[ n \right] \!=\! \frac{{\rho _k^2\left[ {n - \lambda } \right]{p_{{\mathrm{d}}}}N^2{{\left| {\sum\limits_{l = 1}^L {\sqrt {{\mu _{kl}}} {\gamma _{kl}}} } \right|}^2}}}{{{p_{{\mathrm{d}}}}N\!\sum\limits_{i = 1}^K {\!\sum\limits_{l = 1}^L {{\!\mu _{il}}} {\gamma _{il}}{\beta _{kl}}}  \!+\! \rho _k^2\left[ {n \!-\! \lambda } \right]{p_{{\mathrm{d}}}}N^2\!\!\!\sum\limits_{i \in {\mathcal{P}_k}\setminus \{ k \}} {{{\left| {\sum\limits_{l = 1}^L {\!\sqrt {{\mu _{il}}} \sqrt {{\gamma _{kl}}{\gamma _{il}}} } } \right|}^2}}  \!+\! {{\sigma _{\mathrm{d}}^2}}}}.
\end{align}
When $N = 1$ and the effect of channel aging vanishes, formula \eqref{sinr_coh_closed1} will reduce to \cite[Eq. (24)]{Ngo2017Cell}.

\subsection{Non-coherent Downlink Transmission}

In this section, we consider non-coherent joint transmission in downlink CF massive MIMO.
It means each AP is allowed to transmit to each UE but sends a different data symbol than the other APs, to alleviate the phase-synchronization requirements that coherent transmission put on the APs.
Using maximum ratio precoding, the transmitted signal from AP $l$ is
\begin{align}
{{\mathbf{x}}_l}\left[ n \right] =\sqrt{p_{{\text{d}}}} \sum\limits_{i = 1}^K {{{{\mathbf{\hat h}}}_{il}}\left[ \lambda  \right] {\sqrt {{\mu _{il}}} {q_{il}}\left[ n \right]} },
\end{align}
where ${q_{il}}\left[ n \right] \sim \mathcal{C}\mathcal{N}\left( {0,1} \right)$ is the symbol sent to UE $i$ which is different for all APs,
${p_{\text{d}}}$ denotes the downlink transmission power, and ${\mu _{il}}$ is power control coefficients chosen to satisfy the downlink power constraint as $\mathbb{E}\left\{ {{{\left| {{{\mathbf{x}}_l}\left[ n \right]} \right|}^2}} \right\} \leqslant {p_{\text{d}}}$.
With the help of \eqref{h_il}, the received signal at the $k$th UE at time instant $n$ is expressed as
\begin{align}
  &r_k^{{\text{nc}}}\left[ n \right] = \sum\limits_{l = 1}^L {{\mathbf{h}}_{kl}^{\text{H}}\left[ n \right]{{\mathbf{x}}_l}} \left[ n \right] + {w_k}\left[ n \right] = \sqrt{p_{\text{d}}}\sum\limits_{i = 1}^K {\sum\limits_{l = 1}^L {{\mathbf{h}}_{kl}^{\text{H}}\left[ n \right]{{{\mathbf{\hat h}}}_{il}}\left[ \lambda  \right]\sqrt {{\mu _{il}}} {q_{il}}\left[ n \right]} }  + {w_k}\left[ n \right] \notag \\
   &= {\rho _k}\left[ {n - \lambda } \right]\sqrt{p_{\text{d}}}\sum\limits_{l = 1}^L {{\mathbf{h}}_{kl}^{\text{H}}\left[ \lambda  \right]{{{\mathbf{\hat h}}}_{kl}}\left[ \lambda  \right]\sqrt {{\mu _{kl}}} {q_{kl}}\left[ n \right]}  + {{\bar \rho }_k}\left[ {n - \lambda } \right]\sqrt{p_{\text{d}}}\sum\limits_{l = 1}^L {{\mathbf{u}}_{kl}^{\text{H}}\left[ n \right]{{{\mathbf{\hat h}}}_{kl}}\left[ \lambda  \right]\sqrt {{\mu _{kl}}} {q_{kl}}\left[ n \right]} \notag\\
   &+ \sqrt{p_{\text{d}}}\sum\limits_{i \ne k}^K {\sum\limits_{l = 1}^L {{\mathbf{h}}_{kl}^{\text{H}}\left[ n \right]{{{\mathbf{\hat h}}}_{il}}\left[ \lambda  \right]\sqrt {{\mu _{il}}} {q_{il}}\left[ n \right]} }  + {w_k}\left[ n \right].
\end{align}
After receiving the signals sent by all $L$ APs, in order to detect the signals sent by the different APs, the $k$th user needs to use successive interference cancellation.
The specific idea is that the user first detects the signal sent by the first AP, and treats the remaining signal as interference. By analogy, the user detects the signal sent by the $l$th AP, and regards the signal sent from the $\left(l+1\right)$th AP to the $L$th AP as interference, thus detecting the signal $q_{kl}{\left[ n \right]}$.
\begin{thm}
Using successive interference cancellation to detect the non-coherent downlink transmission, the downlink SE of UE $k$ is
\begin{align}\label{SE_c}
{\mathrm{SE}}_k^{{\mathrm{nc}}} = \frac{1}{{{\tau _c}}}\sum\limits_{n=\lambda}^{\tau_c} {{{\log }_2}\left( {1 + {\mathrm{SINR}}_k^{{\mathrm{nc}}}\left[ n \right]} \right)},
\end{align}
where
\begin{align}\label{sinr_nc}
{\mathrm{SINR}}_k^{{\mathrm{nc}}}\!\left[ n \right] \!=\! \frac{{\rho _k^2\left[ {n - \lambda } \right]{p_{\mathrm{d}}}\sum\limits_{l = 1}^L {{\mu _{kl}}{{\left| {\mathbb{E}\left\{ {{\mathbf{h}}_{kl}^{\mathrm{H}}\left[ \lambda  \right]{{{\mathbf{\hat h}}}_{kl}}\left[ \lambda  \right]} \right\}} \right|}^2}} }}{{{p_{\mathrm{d}}}\sum\limits_{i = 1}^K {\sum\limits_{l = 1}^L {{\mu _{il}}\mathbb{E}\left\{ {{{\left| {{\mathbf{h}}_{kl}^{\mathrm{H}}\left[ n \right]{{{\mathbf{\hat h}}}_{il}}\left[ \lambda  \right]} \right|}^2}} \right\} \!-\! \rho _k^2\left[ {n - \lambda } \right]{p_{\mathrm{d}}}\sum\limits_{l = 1}^L {{\mu _{kl}}{{\left| {\mathbb{E}\left\{ {{\mathbf{h}}_{kl}^{\mathrm{H}}\left[ \lambda  \right]{{{\mathbf{\hat h}}}_{kl}}\left[ \lambda  \right]} \right\}} \right|}^2}}  \!+\! \sigma _{\mathrm{d}}^2} } }}.
\end{align}
Computing every term of \eqref{sinr_nc}, ${\mathrm{SINR}}_k^{{\mathrm{coh}}}\left[ n \right] $ can be further expressed in closed form as
\begin{align}
{\mathrm{SINR}}_k^{{\mathrm{nc}}}\left[ n \right] = \frac{{\rho _k^2\left[ {n - \lambda } \right]{p_{\mathrm{d}}}\sum\limits_{l = 1}^L {{\mu _{kl}}} {{\left| {{\mathrm{tr}}\left( {{{\mathbf{Q}}_{kl}}} \right)} \right|}^2}}}{{{p_{\mathrm{d}}}\sum\limits_{i = 1}^K {\sum\limits_{l = 1}^L {{\mu _{il}}} {\mathrm{tr}}\left( {{{\mathbf{Q}}_{il}}{{\mathbf{R}}_{kl}}} \right)}  + \rho _k^2\left[ {n - \lambda } \right]{p_{\mathrm{d}}}\sum\limits_{i \in {\mathcal{P}_k}\setminus \{ k \}}^K {\sum\limits_{l = 1}^L {{\mu _{il}}{{{\left| {{\mathrm{tr}}\left( {{{{\mathbf{\bar Q}}}_{kil}}} \right)} \right|}^2}}}  + \sigma _{\mathrm{d}}^2} }}.
\end{align}

\end{thm}
\begin{IEEEproof}
Please refer to Appendix C.
\end{IEEEproof}

Note that the decoding order between the APs does not matter when it comes to the SE. But the individual signals must be encoded based on a particular choice of decoding order.

\subsection{Downlink Statistical Channel Cooperation Power Control}

We extend the power control policy in \cite{interdonato2019scalability} with a pre-determined function including global statistical channel information and obtain the SCCPC coefficients as
\begin{align}
{\mu _{kl}} = \frac{{\bar \beta _k^{-1} }}{{\sum\nolimits_{i = 1}^K {{\text{tr}}\left( {{{\mathbf{Q}}_{il}}} \right)\bar \beta _i^{-1} } }},k = 1, \ldots K, l = 1, \ldots L,
\end{align}
where
\begin{align}
{{\bar \beta }_k}{\text{ = }}\frac{{\sum\nolimits_{l = 1}^L {{\beta _{kl}}} }}{L}.
\end{align}

\section{Total Energy Efficiency}\label{se:TEE}

In this section, we investigate the total EE of uplink and downlink CF massive MIMO systems taking into account a realistic power consumption model.
\subsection{Power Consumption Model}

Each resource block occupies $\tau_c$ time instants, uplink training occupies $\tau_p$ time instants, uplink and downlink data transmission both occupies $\left( {{\tau _c} - {\tau _p}} \right)$ time instants, and circuit consumption occurs throughout the resource block.
Therefore, the total power consumption of uplink and downlink resource block can be defined \mbox{as \cite{bjornson2015optimal,ngo2017total}}
\begin{align}
{P_{{\text{total}}}} = \frac{{ {\tau_c + \tau _p}}}{{{2\tau _c}}}P_{{\text{TX}}}^{{\text{ul}}} + \frac{{ {\tau_c-\tau _p}}}{{{2\tau _c}}}P_{{\text{TX}}}^{{\text{dl}}} + {P_{{\text{CP}}}},
\end{align}
where $P_{{\text{TX}}}^{{\text{ul}}}$ and $P_{{\text{TX}}}^{{\text{dl}}}$ are respectively the uplink and downlink power amplifiers (PA) due to radiated transmit power and PA dissipation, ${P_{{\text{CP}}}}$ refers to the circuit power (CP) consumption. Furthermore, the uplink and downlink power consumption are respectively given by
\begin{align}
P_{{\text{TX}}}^{{\text{ul}}} &= \sum\limits_{k = 1}^K {{p_{\text{u}}}{\sigma ^2}\frac{{{\eta _k}}}{{{\partial _k}}}} ,\\
P_{{\text{TX}}}^{{\text{dl}}} &= \sum\limits_{l = 1}^L {\frac{1}{{{\partial _l}}}{p_{\text{d}}}{\sigma ^2}\sum\limits_{k = 1}^K {{\mu _{kl}}{\text{tr}}\left( {{{\mathbf{Q}}_{kl}}} \right)} },
\end{align}
where $\partial _k$ is the PA efficiency at UE $k$ and $\partial _l$ is the PA efficiency at AP $l$. The CP consumption ${P_{{\text{CP}}}}$ is obtained as
\begin{align}
{P_{{\text{CP}}}} = \sum\limits_{k = 1}^K {{P_{{\text{ue}},k}}}  + \sum\limits_{l = 1}^L {N{P_{{\text{ap}},l}}}  + \sum\limits_{l = 1}^L {{P_{{\text{bh}},l}}},
\end{align}
where $P_{{\text{ue}},k}$ denotes the required power to run circuit components at UE $k$, $P_{{\text{ap}},l}$ is the internal power required to run the circuit  components related to each antenna of the $l$th AP. The fronthaul power consumption from the $l$th AP to the CPU is obtained as
\begin{align}
{P_{{\text{bh}},l}} = {P_{0,l}} + B \cdot {\text{S}}{{\text{E}}_{{\text{sum}}}} \cdot {P_{{\text{bt}},l}},
\end{align}
where ${P_{0,l}}$ is a fixed power consumption of each fronthaul (traffic-independent power) which may depend on the distances between the APs and the CPU and the system topology, ${P_{{\text{bt}},l}}$ is the traffic-dependent power (in Watt per bit/s), and $B$ is the system bandwidth. $\text{SE}_\text{sum}$ denotes the sum SE over all UEs of uplink and downlink. We choose uplink LSFD and downlink coherent transmission to calculate $\text{SE}_\text{sum}$ as
\begin{align}\label{sum_SE}
{\text{S}}{{\text{E}}_{{\text{sum}}}} = \sum\limits_{k = 1}^K {\left( {\frac{1}{2}{\text{SE}}_k^{{\text{LSFD}}} + \frac{1}{2}{\text{SE}}_k^{{\text{coh}}}} \right)} .
\end{align}
Finally, we can derive the total power consumption as
\begin{align}
{P_{{\text{total}}}} &= \sum\limits_{k = 1}^K {\left( {\frac{{ {\tau_c + \tau _p}}}{{{2\tau _c}}}{p_{\text{u}}}{\sigma ^2}\frac{{{\eta _k}}}{{{\partial _k}}} + {P_{{\text{ue}},k}}} \right)}  + \sum\limits_{l = 1}^L {\left( {\frac{{{\tau_c-\tau _p}}}{{{2\tau _c}}}\frac{1}{{{\partial _l}}}{p_{\text{d}}}{\sigma ^2}\sum\limits_{k = 1}^K {{\mu _{kl}}{\text{tr}}\left( {{{\mathbf{Q}}_{kl}}} \right)}  + N{P_{{\text{ap}},l}} + {P_{0,l}}} \right)}  \notag\\
&+ B\left( {\sum\limits_{l = 1}^L {{P_{{\text{bt}},l}}} } \right){\text{S}}{{\text{E}}_{{\text{sum}}}}.
\end{align}

\subsection{Total Energy Efficiency}
The total EE (bit/Joule) is defined as the sum throughput (bit/s) divided by the total power consumption (Watt) in the network:
\begin{align}\label{tatal_EE}
{\text{E}}{{\text{E}}_{{\text{total}}}} = \frac{{B \cdot {\text{S}}{{\text{E}}_{{\text{sum}}}}}}{{{P_{{\text{total}}}}}}.
\end{align}

\section{Numerical Results}\label{se:NR}

We consider a simulation setup where $L$ APs and $K$ UEs are independently and uniformly distributed within a square of size $0.5 \ {\mathrm{km}}\times0.5\ {\mathrm{km}}$. We utilize the three-slope propagation model from \cite{Ngo2017Cell} as
\begin{align}
{{\beta _{kl}}}\left[ {{\mathrm{dB}}} \right] \!= \!\left\{ {\begin{array}{*{20}{c}}
  { - 81.2,{d_{kl}} < 10{\mathrm{m}}} \\
  { \!\!\!- 61.2 \!- \!20{{\log }_{10}}\left( {\frac{{{d_{kl}}}}{{1{\mathrm{m}}}}} \right),10{\mathrm{m}} \!\leqslant {d_{kl}} < 50{\mathrm{m}}} \\
  { \!- 35.7 \!- \!35{{\log }_{10}}\left( {\frac{{{d_{kl}}}}{{1{\mathrm{m}}}}} \right) \!+ \!{F_{kl}},{d_{kl}} \geqslant 50{\mathrm{m}}},
\end{array}} \right.
\end{align}
where $d_{kl}$ is the horizontal distance between UE $k$ and AP $l$. The shadowing term ${F_{kl}} \sim \mathcal{N}\left( {0,{8^2}} \right)$ only appears when the distance is larger than 50m and the terms are correlated as
\begin{align}
\mathbb{E}\left\{ {{F_{kl}}{F_{ij}}} \right\} = \frac{{{8^2}}}{2}\left( {{2^{ - {\delta _{ki}}/100{\mathrm{m}}}} + {2^{ - {\upsilon _{lj}}/100{\mathrm{m}}}}} \right),
\end{align}
where $\delta _{ki}$ is the distance between UE $k$ and UE $i$, $\upsilon  _{lj}$ is the distance between AP $l$ and AP $j$.
We consider communication at the carrier frequency $f_c=2$ GHz. The pilot transmit power is ${p_1} =  \ldots  = {p_K} = 20$ dBm, the uplink transmission power is $p_\text{u}=20$ dBm, and the downlink transmission power is $p_\text{d}=23$ dBm. The bandwidth is $B=20$ MHz, and the noise power is $\sigma^2=-96$ dBm. The length of one time instant is $T_s=0.01$ ms, and the power consumption parameters as in TABLE \ref{table} \cite{ngo2017total,bashar2019energy}.

\begin{table}[t]
\caption{The Power Consumption Parameters.}
\vspace{4mm}
\centering
\setlength{\tabcolsep}{8mm}{
\begin{tabular}{|c|c|}
    \hline
    \hline
    Parameter & Values \cr\hline
    Power amplifier efficiency at APs and UEs, $\partial_l, \forall l$ and $\partial_k,\forall k$.  & 0.4  \cr\hline
    Internal power component at APs/antennas, $P_{\text{ap},l}, \forall l$ & 0.2 W  \cr\hline
    Internal power component at UEs, $P_{\text{ue},k}, \forall k$& 0.1 W   \cr\hline
    Fixed power consumption of each fronthaul, $P_{0,l}, \forall l$ & 0.825 W  \cr\hline
    Traffic-dependent fronthaul power, $P_{\text{bt},l}, \forall l$ & 0.25 W/Gbit/s  \cr\hline
    \hline
\end{tabular}}
\label{table}
\end{table}

\subsection{The Length of Resource Blocks}
\begin{figure}[t]
\begin{minipage}[t]{0.48\linewidth}	
\centering
\includegraphics[scale=0.55]{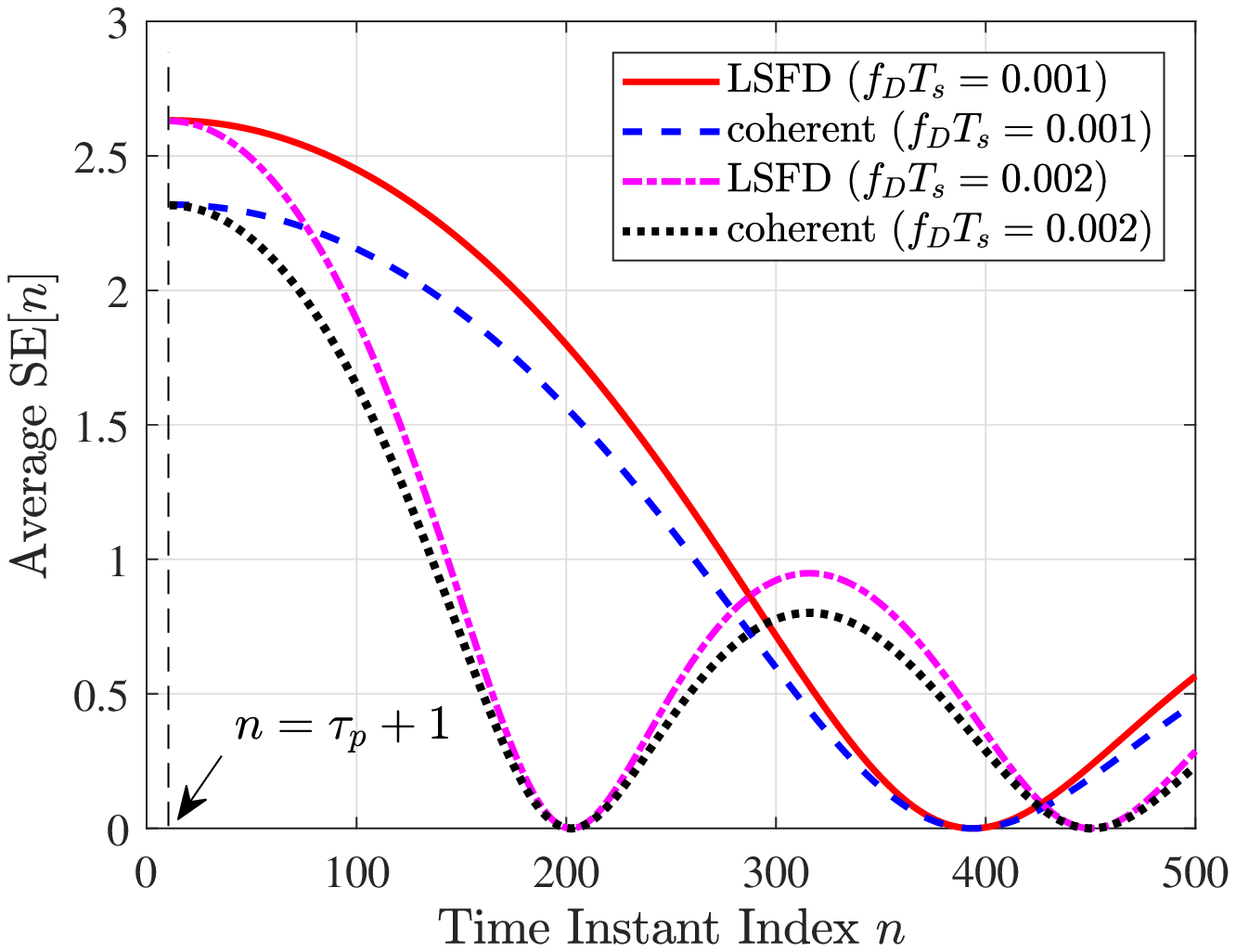}
\caption{Uplink and downlink SE with time instant index.} \vspace{-4mm}
\label{time_instant_index}
\end{minipage}
\hfill
\begin{minipage}[t]{0.48\linewidth}
\centering
\includegraphics[scale=0.55]{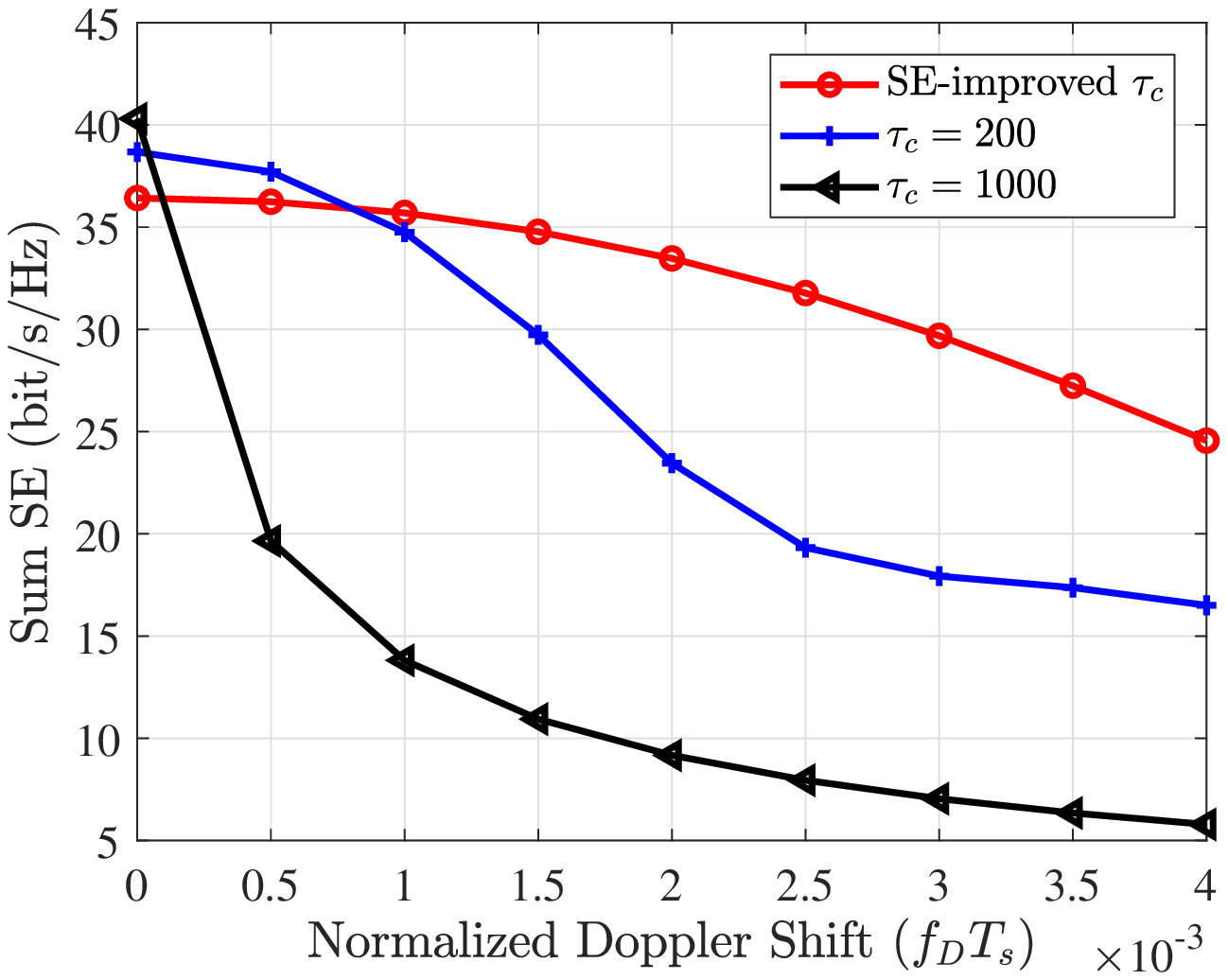}
\caption{Sum SE with the value of $f_DT_s$ under different lengths of the resource block.} \vspace{-4mm}
\label{length_tauc}
\end{minipage}
\end{figure}

In Fig.~\ref{time_instant_index}, we shows the uplink and downlink average ${\text{SE}}\left[ n \right]$ (averaged over all the UEs at the time instant $n$) at first 500 time instant index in an infinitely long resource block, respectively.
The peak of fluctuation is getting smaller and smaller with the time instant index, and the first zero position moves to the left side when increasing the normalized Doppler shift $f_DT_s$. Therefore, we need to design a reasonable length of resource block for reducing the impact from channel aging.
Considering $\tau_p=10$ in Fig.~\ref{length_tauc}, we compare the sum SE (defined in \eqref{sum_SE}) against the value of $f_DT_s$ with different length of the resource block. We observe that the faster the sum SE goes down as the $\tau_c$ increases. In order to reduce the effect of channel aging, we make $\tau_c$ not greater than the value of the first zero of average ${\text{SE}}\left[ n \right]$. Within the range of the Doppler frequency shift, we use the maximum $f_DT_s$ to calculate the value of the first zero, which is larger than the length of resource block. As shown in Fig.~\ref{length_tauc}, simple and practical SE-improved method of determining $\tau_c$ makes sum SE more stable within the considered range of Doppler frequency shift. In the following, we consider $0\leqslant f_DT_s \leqslant 0.002$ and $\tau_c=200$.

\subsection{Spectral Efficiency and Total Energy Efficiency Analysis}
\begin{figure}[t]
\begin{minipage}[t]{0.48\linewidth}	
\centering
\includegraphics[scale=0.55]{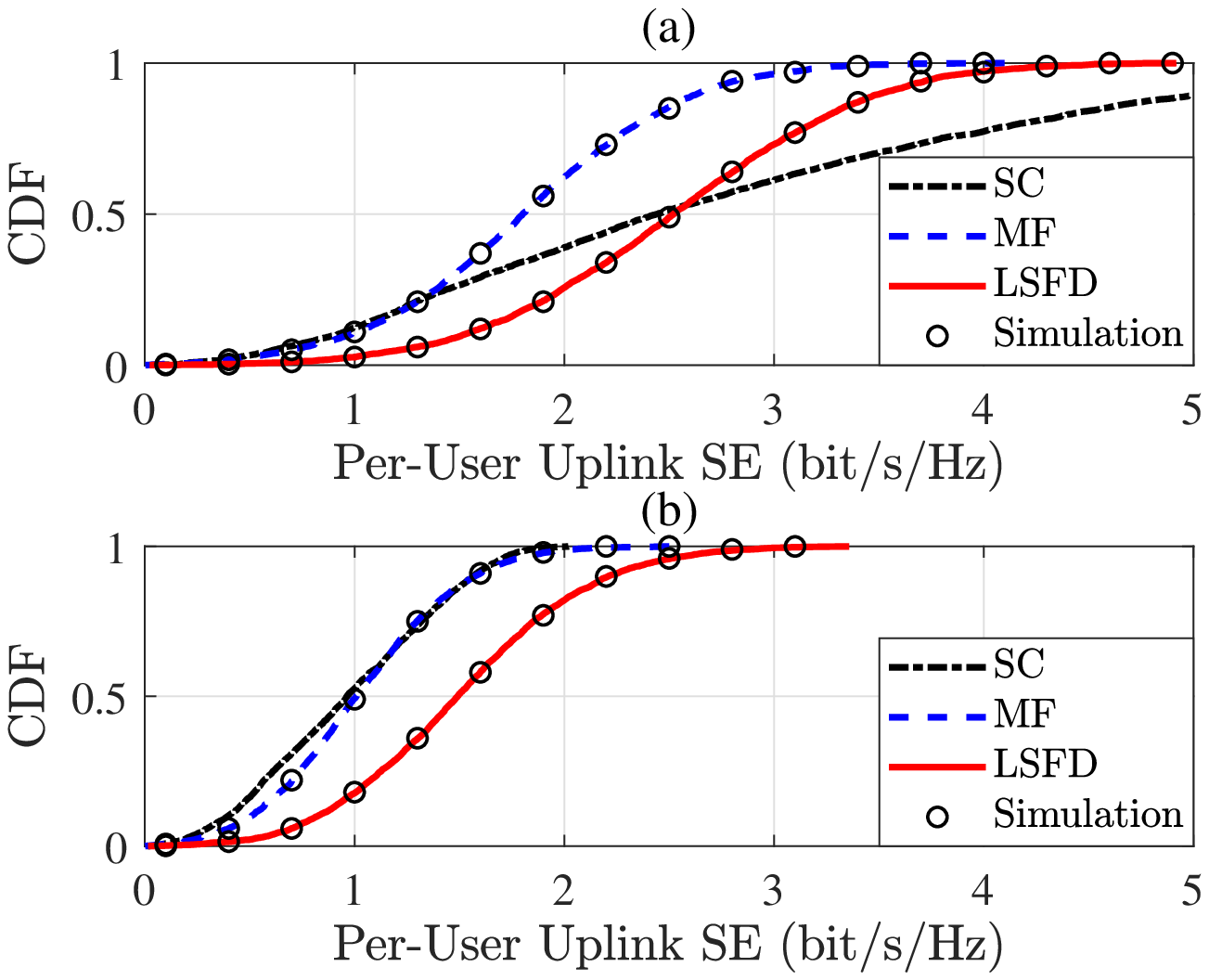}
\caption{CDF of per-user uplink SE of CF and SC systems with full power ($L=100, K=20, N=2, \text{ASD}=30^\text{o}, \tau_p=10$). (a) $f_DT_s=0$; (b) $f_DT_s=0.002$.} \vspace{-4mm}
\label{up_CDF_SE}
\end{minipage}
\hfill
\begin{minipage}[t]{0.48\linewidth}
\centering
\includegraphics[scale=0.55]{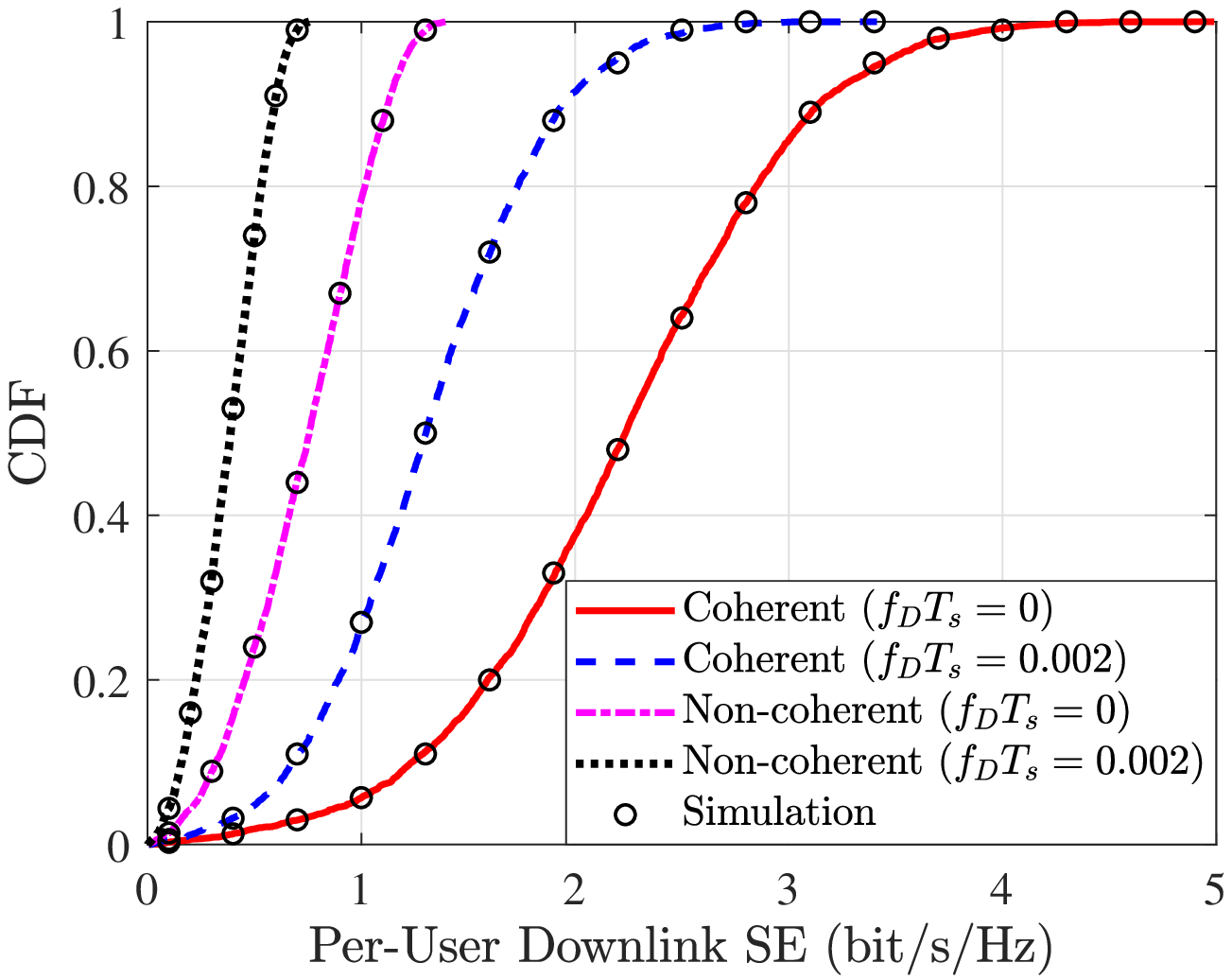}
\caption{CDF of per-user downlink SE of coherent and non-coherent transmission with full power ($L=100, K=20, N=2, \text{ASD}=30^\text{o}, \tau_p=10$).} \vspace{-4mm}
\label{down_CDF_SE}
\end{minipage}
\end{figure}

Fig.~\ref{up_CDF_SE} compares the CDF of the per-user uplink SE achieved in the CF massive MIMO and SC systems with full power with $f_dT_s\!=\!0$ and $0.002$, respectively. The randomness is due to the random AP and UE locations.
It is clear that the LSFD system performs better than MF and SC systems at the median and 95\%-likely SE points.
Increasing the normalized Doppler shift $f_DT_s$ from 0 to 0.002 causes $41\%$ median SE loss of LSFD, $44\%$ median SE loss of MF and $60\%$ median SE loss of SC, respectively. The reason is that LSFD utilizes the knowledge of the fading statistics in the entire network to calculate weight coefficient and thereby mitigate interference, which can effectively counter the impact of channel aging.
Note that the SC system makes use of a tighter capacity bound that utilizes the channel estimates in the data detection but anyway performs poorly under channel aging.

Fig.~\ref{down_CDF_SE} shows the CDF of the per-user downlink SE for coherent and non-coherent transmission with full power with $f_dT_s\!=\!0$ and $0.002$,  respectively. These figures demonstrate that the coherent transmission provides substantially higher SE than the non-cohernt transmission whether the UEs are stationary or mobile. Increasing the normalized Doppler shift $f_DT_s$ from 0 to 0.002 causes $42\%$ median SE loss of coherent transmission and $49\%$ median SE loss of non-coherent transmission, respectively. The reason is that the UE in non-coherent transmission detects the signal from APs one by one, which is sensitive to the effect of channel aging.

\begin{figure}[t]
\begin{minipage}[t]{0.48\linewidth}	
\centering
\includegraphics[scale=0.55]{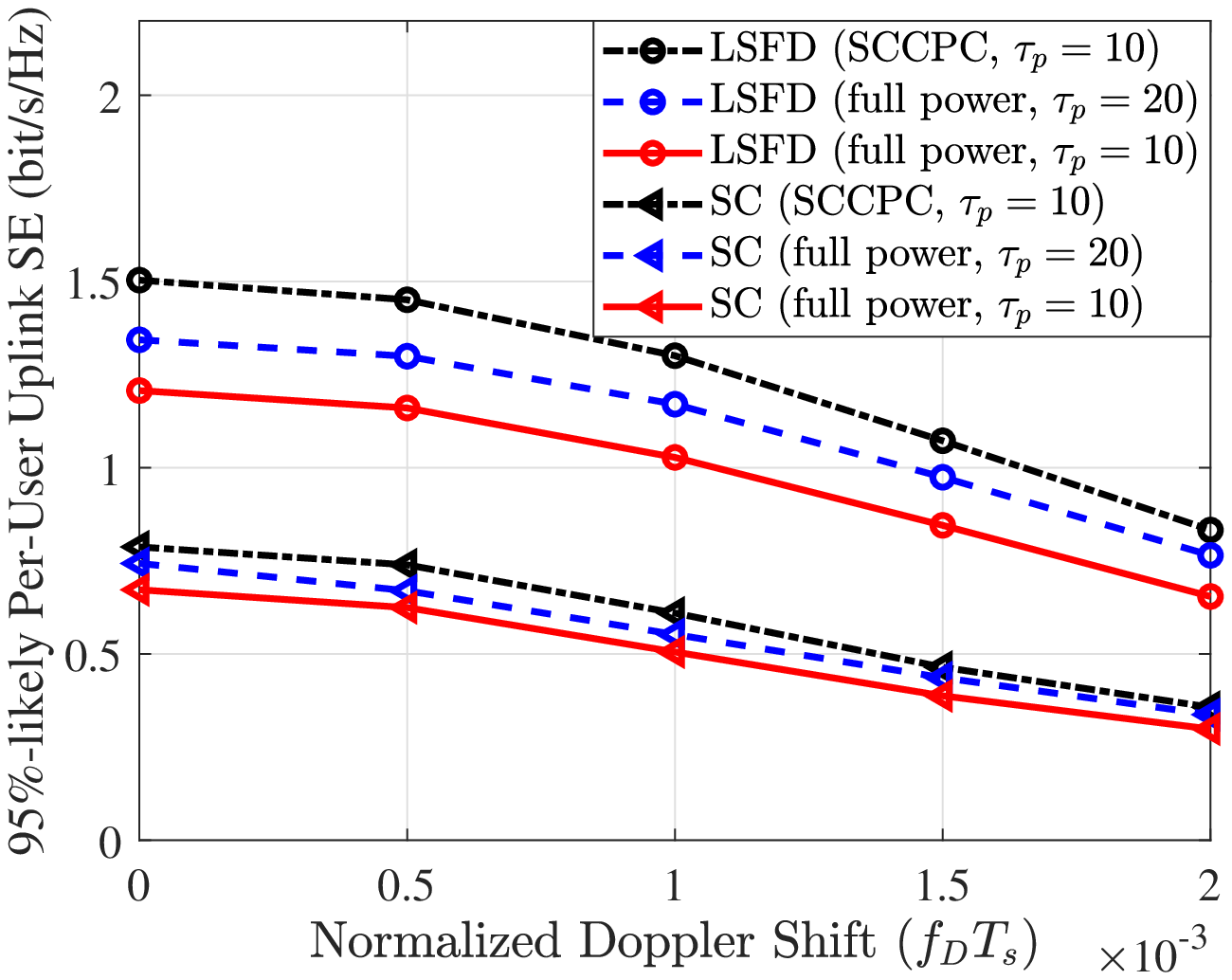}
\caption{95\%-likely per-user uplink SE against the value of $f_DT_s$ for CF and SC systems ($L=100$, $K=20$, $N=2$, $\text{ASD}=30^\text{o}$).} \vspace{-4mm}
\label{up_SE_ft}
\end{minipage}
\hfill
\begin{minipage}[t]{0.48\linewidth}
\centering
\includegraphics[scale=0.55]{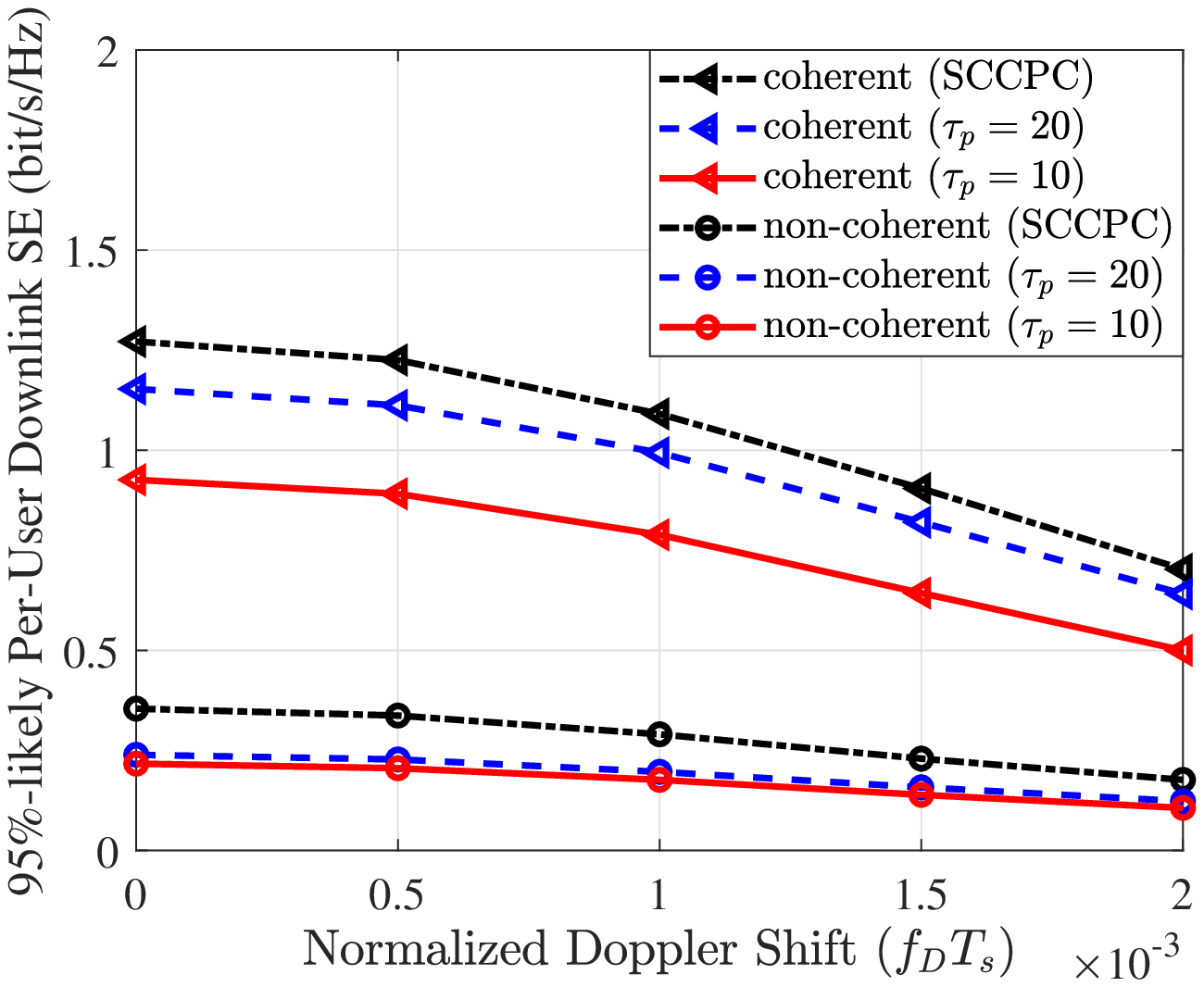}
\caption{95\%-likely per-user downlink SE against the value of $f_DT_s$ for coherent and non-coherent transmission ($L=100$, $K=20$, $N=2$, $\text{ASD}=30^\text{o}$).} \vspace{-4mm}
\label{down_SE_ft}
\end{minipage}
\end{figure}

The 95\%-likely per-user uplink SE with LSFD and of the SC system is shown in Fig.~\ref{up_SE_ft}, as a decreasing function of the normalized Doppler shift $f_DT_s$. We notice that CF with LSFD achieves larger 95\%-likely SE than the corresponding SC system in both low- and high-mobility conditions. Therefore, CF massive MIMO systems are more suitable for mobility scenarios than SC systems.
For both types of systems, the 95\%-likely SE with SCCPC is getting closer to the 95\%-likely SE with full power when $f_DT_s$ varies from 0 to 0.002, especially for SC systems.
The reason is that the self-interference caused by the channel aging effect becomes more dominant than the inter-user interference in high-mobility scenarios, thus the interference reduction due to SCCPC becomes less influential.
Furthermore, the large-scale fading coefficients available at the CPU can be efficiently combined for the power control in CF massive MIMO systems.
The figure also shows results for the case when the length of the training phase $\tau_p$ is increased, which leads to better SE.

Fig.~\ref{down_SE_ft} shows the 95\%-likely per-user downlink SE for coherent and non-coherent transmission is a decreasing function of the normalized Doppler shift $f_DT_s$. Coherent transmission has at least four times 95\%-likely per-user SE than non-coherent transmission whether low- or high-mobility conditions. Compared to non-coherent transmission, using SCCPC in coherent transmission can provide larger 95\%-likely per-user SE.
It is worth noting that increasing $\tau_p$ respectively leads to 25\% and 10\% gain at 95\%-likely per-user SE point of coherent and non-coherent transmission when there is no normalized Doppler shift effect.
Therefore, even if non-coherent transmission is helpful to solve the phase-synchronization problem, coherent transmission is still the better choice.

\begin{figure}[t]
\begin{minipage}[t]{0.48\linewidth}	
\centering
\includegraphics[scale=0.55]{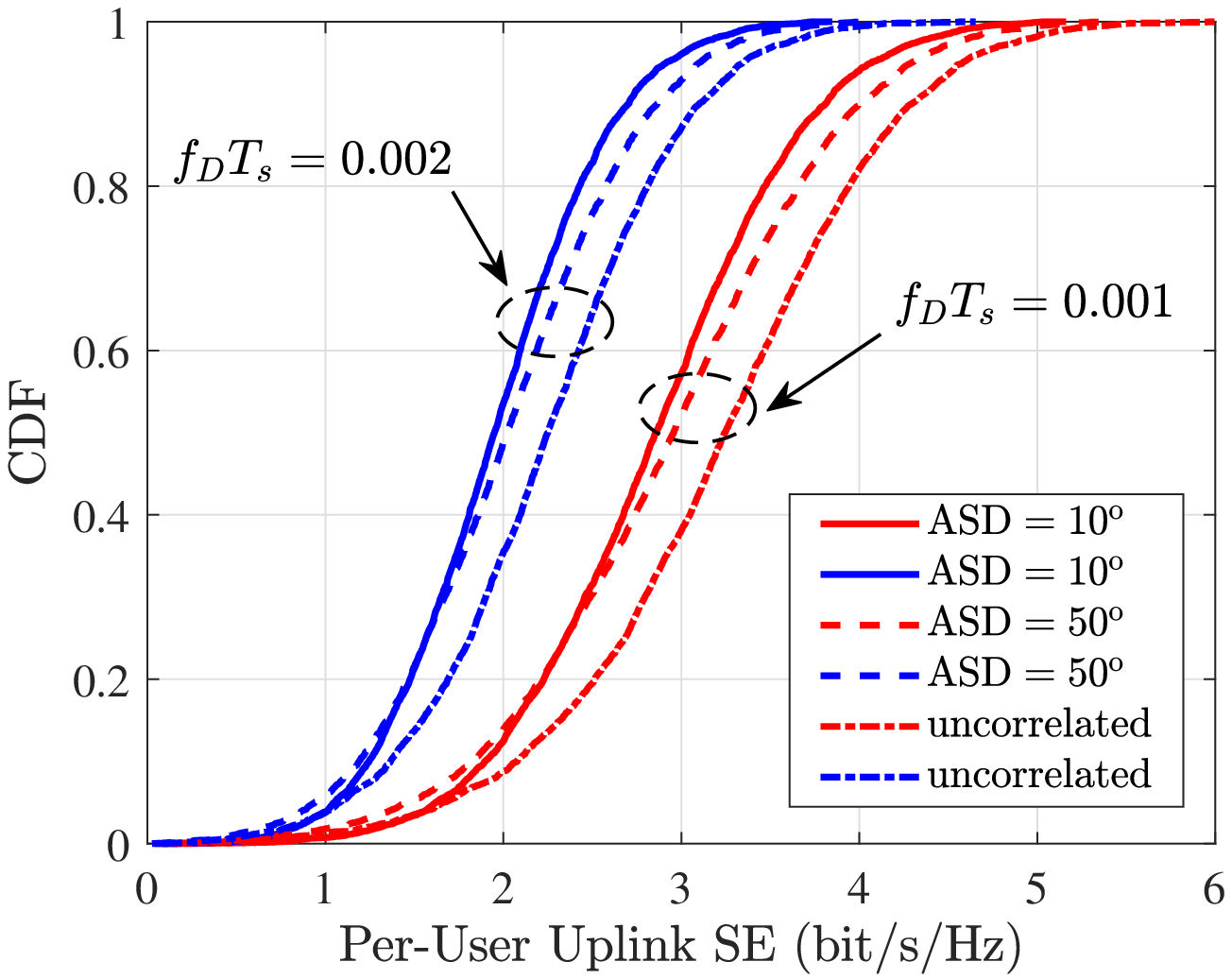}
\caption{CDF of per-user uplink SE of LSFD system with full power ($L=100$, $K=20$, $N=4$, $\tau_p=10$).} \vspace{-4mm}
\label{LSFD_ASD}
\end{minipage}
\hfill
\begin{minipage}[t]{0.48\linewidth}
\centering
\includegraphics[scale=0.55]{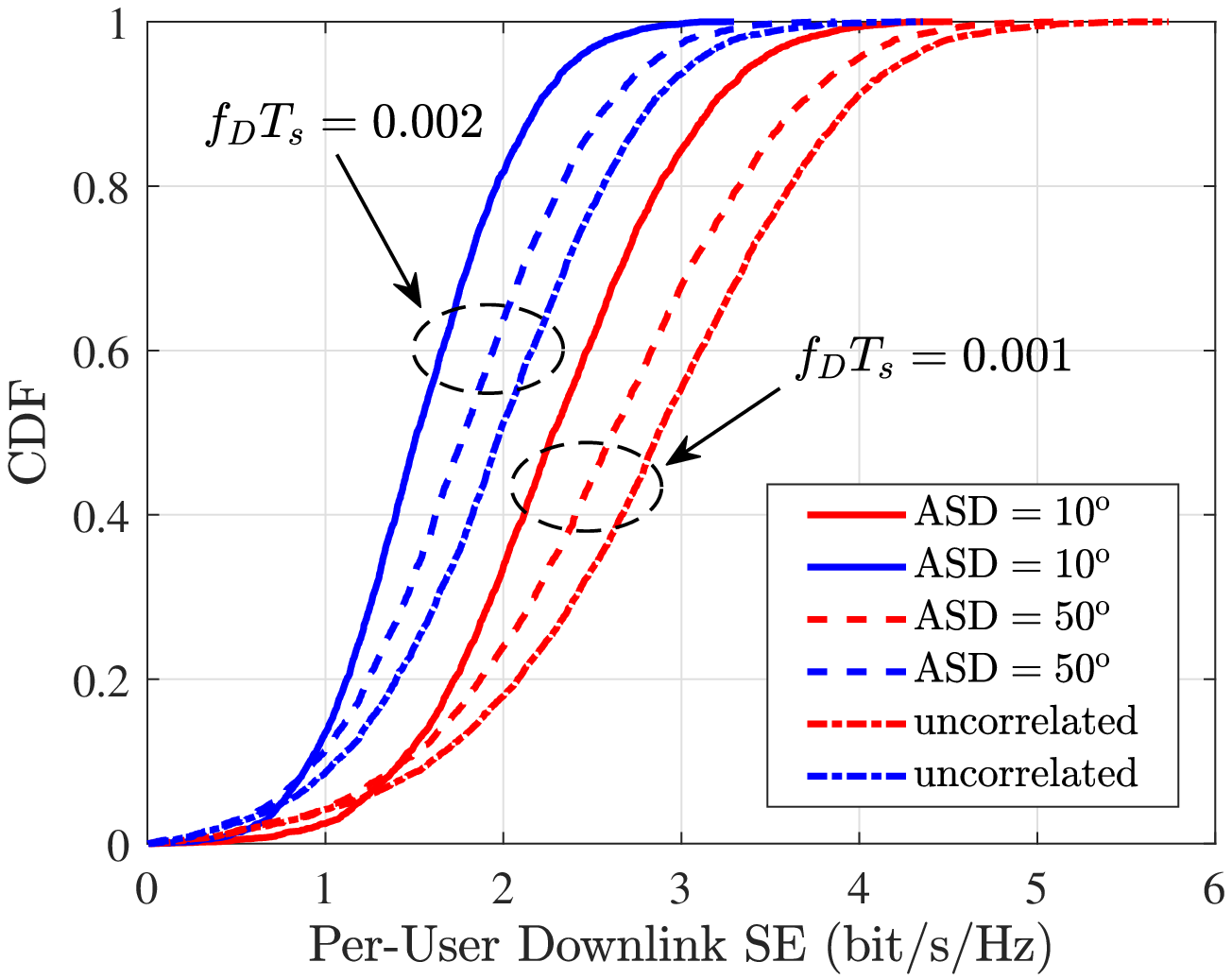}
\caption{CDF of per-user downlink SE of coherent transmission system with full power ($L=100$, $K=20$, $N=4$, $\tau_p=10$).} \vspace{-4mm}
\label{coherent_ASD}
\end{minipage}
\end{figure}

Fig.~\ref{LSFD_ASD} compares the uplink SE differences between low- and high-mobility conditions for LSFD system under different channel correlations.
$\text{ASD}={10^{\text{o}}}$ denotes strong spatial correlation, $\text{ASD}={50^{\text{o}}}$ denotes weak spatial correlation, and uncorrelated denotes there is no spatial correlation. It can be seen that stronger spatial correlation leads to lower SE. When $f_DT_s$ varies from 0.002 to 0.001, the loss in median SE for strong spatial correlation and uncorrelated fading are $0.9$ bit/s/Hz and $1$ bit/s/Hz, respectively. Therefore, larger spatial correlation can reduce the effects of channel aging.
In Fig.~\ref{coherent_ASD}, we consider the same setting as in Fig.~\ref{LSFD_ASD} but with downlink coherent transmission. The effect of channel aging is small in strong spatial correlated channel as expected. Quantitatively, it is 0.7 bit/s/Hz median SE loss for strong spatial correlation and 0.9 bit/s/Hz median SE loss for uncorrelated channel.
It is worth noting that strong spatial correlation is beneficial to poor UEs both in uplink and downlink data transmission.

\begin{figure}[t]
\begin{minipage}[t]{0.48\linewidth}	
\centering
\includegraphics[scale=0.55]{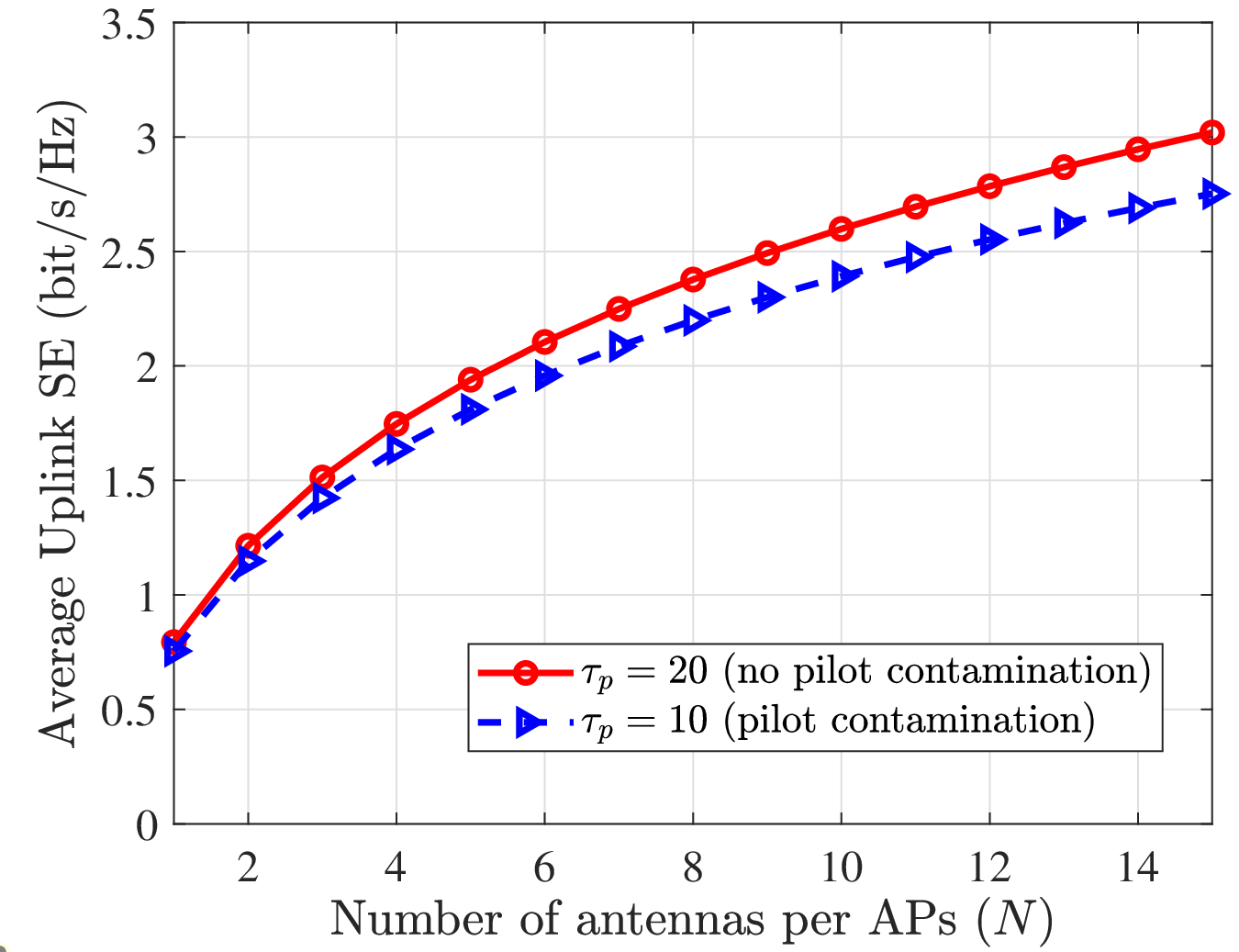}
\caption{SE versus antenna numbers per APs ($L\!=\!100$, $K\!=\!20$, $f_DT_s\!=\!0.002$, uncorrelated).} \vspace{-4mm}
\label{SE_N}
\end{minipage}
\hfill
\begin{minipage}[t]{0.48\linewidth}
\centering
\includegraphics[scale=0.55]{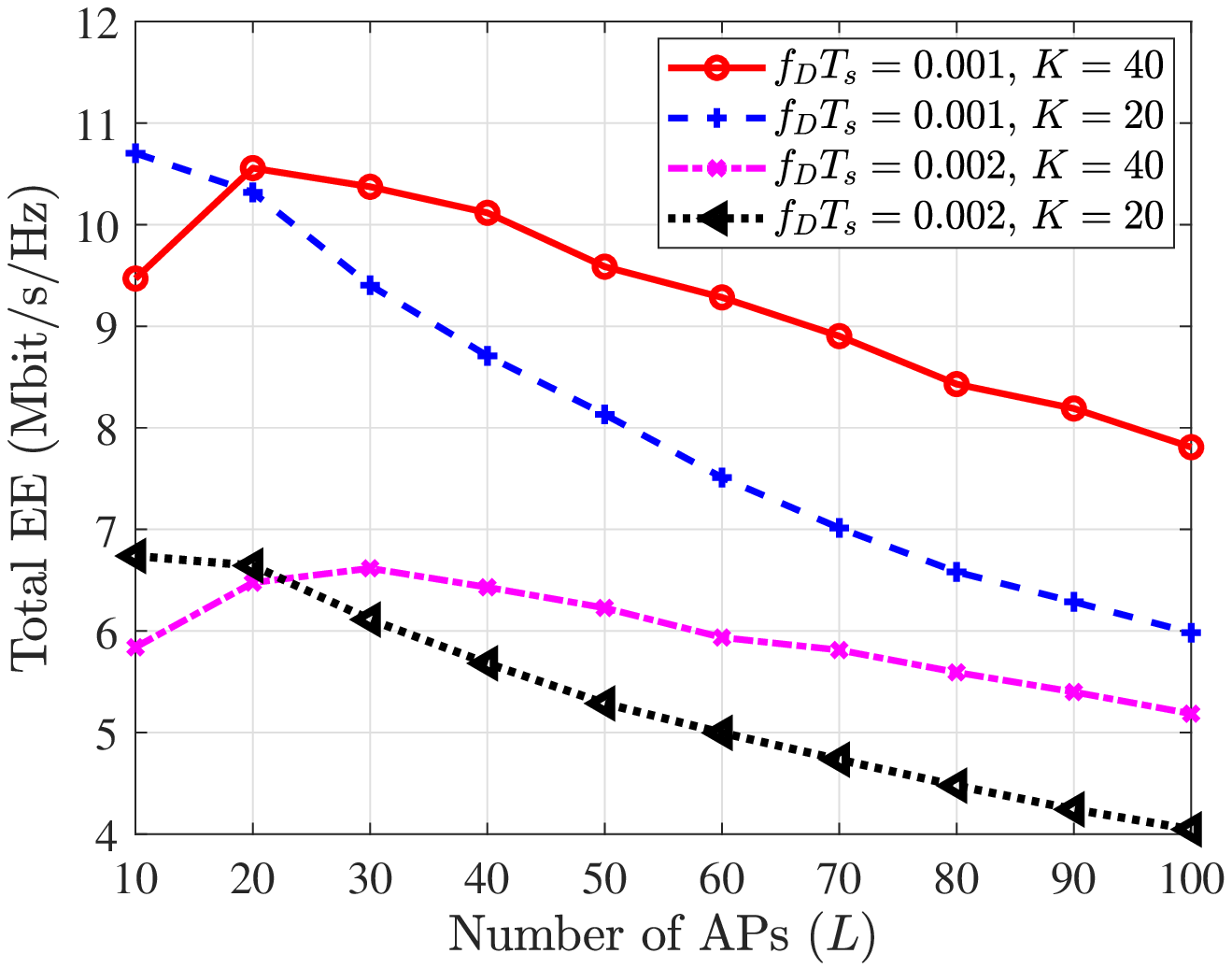}
\caption{Total EE versus the number of APs ($N=2$, $\text{ASD}=10^\text{o}$, $\tau_p=10$).} \vspace{-4mm}
\label{total_EE}
\end{minipage}
\end{figure}

Fig.~\ref{SE_N} shows the average uplink SE versus the number of antennas per AP on the basis of Remark \ref{Re_SE_N}.
When there is no pilot contamination ($\dot b = 0$), the SE keeps growing with the number of antennas. We also find that the SE with pilot contamination grows almost as if there is no pilot contamination.
The reason is that, in systems with few antennas, the pilot-contaminated interference is typically small compared to other types of interference. It is only in cellular massive MIMO systems with very many antennas per AP that pilot contamination can be a major limiting factor. Hence, the expected effect of increasing $N$ in CF massive MIMO is that the SINR grows nearly proportionally to it.

Fig.~\ref{total_EE} shows the total EE in \eqref{total_EE} against the number of APs for different values of $f_DT_s$ and $K$. It is clear that the total EE first increases and then decreases with the increasing of the number of APs $L$. There is an optimal number of APs to achieves the maximum EE. The reason is that the total energy consumption increases linearly with the number of APs, but the SE increases logarithmically. Moreover, having a larger number of UEs can increase the optimal number of APs at the point where we obtain maximum EE, and make the total EE decrease more slowly on the right side of the maximum EE. For example, when $L$ increases from 30 to 100, the total EE for $f_DT_s=0.001$, $K=20$ has a 36\% loss, and the total EE for $f_DT_s=0.001$, $K=40$ has a 25\% loss.
Furthermore, larger normalized Doppler shift $f_DT_s$ leads to a smaller total EE and leads to more APs are preferred for the optimal operating point of EE. Because channel aging reduces SE, which means more antennas are needed to restore balance.
For the case $K=20$, when $f_DT_s$ varies from 0.001 to 0.002, the total EE for $L=10$ has a 37\% loss, and the total EE for $L=100$ has a 32\% loss. Therefore, increasing the number of APs can reduce the impact of channel aging on the total EE.

\section{Conclusions}\label{se:CON}
In this paper, we investigate the uplink and downlink performance of CF massive MIMO, taking into account the impact of channel aging, spatial correlation and pilot contamination. In the uplink, we consider LSFD and MF receiver cooperation, and uncooperative SC system is analysed for comparison. In the downlink, we study the performance of both coherent and non-coherent transmission modes. Based on the channel estimates, we derive novel and exact closed-form expressions for the uplink and downlink SE of the considered system and quantify the channel aging effect. It is important that the channel aging effect degrades the performance of the considered systems, but the CF massive MIMO systems is less affected by SC systems in mobile scenarios and coherent transmission has a higher performance than non-coherent transmission. Furthermore, a practical SCCPC is proposed to improve the SE performance, while the gain from SCCPC gradually reduce as the channel aging becomes stronger. However, multiple antennas and enough pilots can mitigate the impact of channel aging. Weak spatial correlation achieves larger SE, but it is more easily affected by channel aging. We also find that increasing channel aging significantly reduces the EE and leads to more APs are preferred for the optimal operating point of EE. Finally, a method to design the SE-improved length of resource block is provided for a uniform performance under channel aging.

In a multi-carrier system, there will also be correlation between the narrowband subcarriers. If we choose to not make use of that correlation in the algorithmic design, we can treat each subcarrier independently and apply the methods that we have developed separately on each of them. However, in future work, one can investigate how the correlation between subcarriers over the frequency domain can be utilized for improved operation. Some related work on that is \cite{8716688,kashyap2016frequency} and references therein.

\begin{appendices}
\section{Proof of Theorem 1}

We consider a set of $\left(\tau_c-\tau_p\right)$ channel codes, each applied to the $n$th time instant in every resource block, for $n = \lambda , \ldots ,{\tau _c}$.
Using the use-and-then-forget capacity bound in \cite{bjornson2017massive} at every time instant and taking the average, the SE of UE $k$ is
\begin{align}
{\text{S}}{{\text{E}}_k} = \frac{1}{{{\tau _c}}}\sum\limits_{n = \lambda}^{\tau_c} {{{\log }_2}\left( {1 + {\text{SIN}}{{\text{R}}_k}\left[ n \right]} \right)} ,
\end{align}
where
\begin{align}
{{\text{SIN}}{{\text{R}}_k}\left[ n \right]}=\frac{{\mathbb{E}\left\{ {{{\left| {{\text{D}}{{\text{S}}_{k,n}}} \right|}^2}}\right\}}}{{\mathbb{E}\left\{ {{{\left| {{\text{B}}{{\text{U}}_{k,n}}} \right|}^2}} \right\} + \mathbb{E}\!\left\{ {{{\left| {{\text{C}}{{\text{A}}_{k,n}}} \right|}^2}} \right\} +  \sum\limits_{i \ne k}^K {\mathbb{E}\left\{ {{{\left| {{\text{U}}{{\text{I}}_{ki,n}}} \right|}^2}} \right\}}  +  \mathbb{E}\left\{ {{{\left| {{\text{N}}{{\text{S}}_{k,n}}} \right|}^2}} \right\}}}. \notag
\end{align}
We will compute every term of ${{\text{SIN}}{{\text{R}}_k}\left[ n \right]}$ to obtain \eqref{SINR_k}.

\emph{1) Compute $\mathbb{E}\left\{ {{\mathbf{\hat h}}_{kl}^{\text{H}}\left[ \lambda  \right]{{\mathbf{h}}_{il}}\left[ \lambda  \right]} \right\}$:}
Based on the properties of MMSE estimation, ${{{\mathbf{\hat h}}}_{il}}\left[ \lambda  \right]$ and ${{{\mathbf{\tilde h}}}_{il}}\left[ \lambda  \right]$ are independent \cite{bjornson2017massive}.
When $i \in \mathcal{P}_k$, ${\mathbf{\hat h}}_{kl}\left[ \lambda  \right]$ is correlated with ${{\mathbf{h}}_{il}}\left[ \lambda  \right]$. Utilizing \eqref{hhat}, we have
\begin{align}
  \mathbb{E}\left\{ {{\mathbf{\hat h}}_{kl}^{\text{H}}\left[ \lambda  \right]{{\mathbf{h}}_{il}}\left[ \lambda  \right]} \right\} &=\mathbb{E}\left\{ {{\mathbf{\hat h}}_{kl}^{\text{H}}\left[ \lambda  \right]\left( {{{{\mathbf{\hat h}}}_{il}}\left[ \lambda  \right] + {{{\mathbf{\tilde h}}}_{il}}\left[ \lambda  \right]} \right)} \right\}= {\text{tr}}\left( {\mathbb{E}\left\{ {{{\mathbf{\hat h}}_{il}}\left[ \lambda  \right]{\mathbf{\hat h}}_{kl}^{\text{H}}\left[ \lambda  \right]} \right\}} \right) \notag \\
   &= {\text{tr}}\left( {\mathbb{E}\left\{ {{\rho _k}\left[ {\lambda  - {t_k}} \right]\sqrt {{p_k}} {\rho _i}\left[ {\lambda  - {t_k}} \right]\sqrt {{p_i}} {{\mathbf{R}}_{il}}{{\mathbf{\Psi }}_{kl}}{{\mathbf{z}}_l}\left[ {{t_k}} \right]{\mathbf{z}}_l^{\text{H}}\left[ {{t_k}} \right]{{\mathbf{\Psi }}_{kl}}{{\mathbf{R}}_{kl}}} \right\}} \right) \notag \\
   &= {\rho _k}\left[ {\lambda  - {t_k}} \right]\sqrt {{p_k}} {\rho _i}\left[ {\lambda  - {t_k}} \right]\sqrt {{p_i}} {\text{tr}}\left( {{{\mathbf{R}}_{il}}{{\mathbf{\Psi }}_{kl}}{{\mathbf{R}}_{kl}}} \right).
\end{align}
When $i \notin {\mathcal{P}_k}$, ${\mathbf{\hat h}}_{kl}\left[ \lambda  \right]$ and ${\mathbf{\hat h}}_{il}\left[ \lambda  \right]$ are independent, thereby we have $\mathbb{E}\left\{ {{\mathbf{\hat h}}_{kl}^{\text{H}}\left[ \lambda  \right]{{\mathbf{h}}_{il}}\left[ \lambda  \right]} \right\} = 0$. With the help of \eqref{Q_hat}, we can obtain
\begin{align}\label{E1}
\mathbb{E}\left\{ {{\mathbf{\hat h}}_{kl}^{\text{H}}\left[ \lambda  \right]{{\mathbf{h}}_{il}}\left[ \lambda  \right]} \right\} = \left\{ {\begin{array}{*{20}{c}}
  {{{\text{tr}}\left( {{{{\mathbf{\bar Q}}}_{kil}}} \right)} ,i \in {\mathcal{P}_k}} \\
  {0,i \notin {\mathcal{P}_k}.}
\end{array}} \right.
\end{align}

\emph{2) Compute $\mathbb{E}\left\{ {{{\left| {{\mathbf{\hat h}}_{kl}^{\text{H}}\left[ \lambda  \right]{{\mathbf{h}}_{il}}\left[ \lambda  \right]} \right|}^2}} \right\}$:}
When $i \in {\mathcal{P}_k}$, we have
\begin{align}\label{S1}
  \mathbb{E}\left\{ {{{\left| {{\mathbf{\hat h}}_{kl}^{\text{H}}\left[ \lambda  \right]{{\mathbf{h}}_{il}}\left[ \lambda  \right]} \right|}^2}} \right\} &= \mathbb{E}\left\{ {{\mathbf{\hat h}}_{kl}^{\text{H}}\left[ \lambda  \right]\left( {{{{\mathbf{\hat h}}}_{il}}\left[ \lambda  \right] + {{{\mathbf{\tilde h}}}_{il}}\left[ \lambda  \right]} \right)\left( {{\mathbf{\hat h}}_{il}^{\text{H}}\left[ \lambda  \right] + {\mathbf{\tilde h}}_{il}^{\text{H}}\left[ \lambda  \right]} \right){{{\mathbf{\hat h}}}_{kl}}\left[ \lambda  \right]} \right\} \notag \\
  & \!\!\!\!\!= \underbrace {\mathbb{E}\left\{ {{\mathbf{\hat h}}_{kl}^{\text{H}}\left[ \lambda  \right]{{{\mathbf{\hat h}}}_{il}}\left[ \lambda  \right]{\mathbf{\hat h}}_{il}^{\text{H}}\left[ \lambda  \right]{{{\mathbf{\hat h}}}_{kl}}\left[ \lambda  \right]} \right\}}_{{\Upsilon _1}} + \underbrace {\mathbb{E}\left\{ {{\mathbf{\hat h}}_{kl}^{\text{H}}\left[ \lambda  \right]{{{\mathbf{\tilde h}}}_{il}}\left[ \lambda  \right]{\mathbf{\tilde h}}_{il}^{\text{H}}\left[ \lambda  \right]{{{\mathbf{\hat h}}}_{kl}}\left[ \lambda  \right]} \right\}}_{{\Upsilon _2}} .
\end{align}
Plugging \eqref{hhat} into ${{\Upsilon _1}}$, we obtain
\begin{align}
 {{\Upsilon _1}}&= \rho _i^2\left[ {\lambda  - {t_k}} \right]{p_i}\rho _k^2\left[ {\lambda  - {t_k}} \right]{p_k}\mathbb{E}\left\{ {{{\left| {{{\left( {{{\mathbf{R}}_{il}}{{\mathbf{\Psi }}_{kl}}{{\mathbf{z}}_l}\left[ {{t_k}} \right]} \right)}^{\text{H}}}{{\mathbf{R}}_{kl}}{{\mathbf{\Psi }}_{kl}}{{\mathbf{z}}_l}\left[ {{t_k}} \right]} \right|}^2}} \right\} \notag \\
   &= \rho _i^2\left[ {\lambda  - {t_k}} \right]{p_i}\rho _k^2\left[ {\lambda  - {t_k}} \right]{p_k}\mathbb{E}\left\{ {{{\left| {{\mathbf{z}}_l^{\text{H}}\left[ {{t_k}} \right]{{\mathbf{\Psi }}_{kl}}{{\mathbf{R}}_{il}}{{\mathbf{R}}_{kl}}{{\mathbf{\Psi }}_{kl}}{{\mathbf{z}}_l}\left[ {{t_k}} \right]} \right|}^2}} \right\}.
\end{align}
With the help of \cite[Lemma B.14]{bjornson2017massive}, we further obtain
\begin{align}\label{S2}
 {{\Upsilon _1}}&= \rho _i^2\left[ {\lambda  - {t_k}} \right]{p_i}\rho _k^2\left[ {\lambda  - {t_k}} \right]{p_k}\left( {{{\left| {{\text{tr}}\left( {{{\mathbf{\Psi }}_{kl}}{{\mathbf{R}}_{il}}{{\mathbf{R}}_{kl}}{{\mathbf{\Psi }}_{kl}}{\mathbf{\Psi }}_{kl}^{ - 1}} \right)} \right|}^2}} \right. \notag\\
 &\left.{ + {\text{tr}}\left( {{{\mathbf{\Psi }}_{kl}}{{\mathbf{R}}_{il}}{{\mathbf{R}}_{kl}}{{\mathbf{\Psi }}_{kl}}{\mathbf{\Psi }}_{kl}^{ - 1}{{\left( {{{\mathbf{\Psi }}_{kl}}{{\mathbf{R}}_{il}}{{\mathbf{R}}_{kl}}{{\mathbf{\Psi }}_{kl}}} \right)}^{\text{H}}}{\mathbf{\Psi }}_{kl}^{ - 1}} \right)} \right)
 = {\left| {{\text{tr}}\left( {{{{\mathbf{\bar Q}}}_{kil}}} \right)} \right|^2} + {\text{tr}}\left( {{{\mathbf{Q}}_{kl}}{{\mathbf{Q}}_{il}}} \right).
\end{align}
Consequently, we can derive $ {{\Upsilon _2}}$ as
\begin{align}\label{S3}
 {{\Upsilon _2}}= {\text{tr}}\left( {\mathbb{E}\left\{ {{{{\mathbf{\hat h}}}_{kl}}\left[ \lambda  \right]{\mathbf{\hat h}}_{kl}^{\text{H}}\left[ \lambda  \right]} \right\}\mathbb{E}\left\{ {{{{\mathbf{\tilde h}}}_{il}}\left[ \lambda  \right]{\mathbf{\tilde h}}_{il}^{\text{H}}}\left[ \lambda  \right] \right\}} \right) = {\text{tr}}\left( {{{\mathbf{Q}}_{kl}}\left( {{{\mathbf{R}}_{il}} - {{\mathbf{Q}}_{il}}} \right)} \right).
\end{align}
In addition, when $i \notin {\mathcal{P}_k}$, we have
\begin{align}\label{S4}
  \mathbb{E}\left\{ {{{\left| {{\mathbf{\hat h}}_{kl}^{\text{H}}\left[ \lambda  \right]{{\mathbf{h}}_{il}}\left[ \lambda  \right]} \right|}^2}} \right\} &= \mathbb{E}\left\{ {{\mathbf{\hat h}}_{kl}^{\text{H}}\left[ \lambda  \right]{{\mathbf{h}}_{il}}\left[ \lambda  \right]{\mathbf{h}}_{il}^{\text{H}}\left[ \lambda  \right]{{{\mathbf{\hat h}}}_{kl}}\left[ \lambda  \right]} \right\} = {\text{tr}}\left( {\mathbb{E}\left\{ {{\mathbf{\hat h}}_{kl}^{\text{H}}\left[ \lambda  \right]{{\mathbf{h}}_{il}}\left[ \lambda  \right]{\mathbf{h}}_{il}^{\text{H}}\left[ \lambda  \right]{{{\mathbf{\hat h}}}_{kl}}\left[ \lambda  \right]} \right\}} \right) \notag \\
   &= {\text{tr}}\left( {\rho _k^2\left[ {\lambda  - {t_k}} \right]{p_k}{{\mathbf{R}}_{kl}}{{\mathbf{\Psi }}_{kl}}{{\mathbf{R}}_{kl}}{{\mathbf{R}}_{il}}} \right) = {\text{tr}}\left( {{{\mathbf{Q}}_{kl}}{{\mathbf{R}}_{il}}} \right).
\end{align}
By utilizing \eqref{S1}, \eqref{S2}, \eqref{S3} and \eqref{S4}, we obtain
\begin{align}\label{E2}
\mathbb{E}\left\{ {{{\left| {{\mathbf{\hat h}}_{kl}^{\text{H}}\left[ \lambda  \right]{{\mathbf{h}}_{il}}\left[ \lambda  \right]} \right|}^2}} \right\} = {\text{tr}}\left( {{{\mathbf{Q}}_{kl}}{{\mathbf{R}}_{il}}} \right) + \left\{ {\begin{array}{*{20}{c}}
  {{\left| {{\text{tr}}\left( {{{{\mathbf{\bar Q}}}_{kil}}} \right)} \right|^2},i \in {\mathcal{P}_k}} \\
  {0,i \notin {\mathcal{P}_k}}
\end{array}} \right..
\end{align}

\emph{3) Compute $\mathbb{E}\left\{{{{\left| {{\mathrm{D}}{{\mathrm{S}}_{k,n}}} \right|}^2}}\right\}$:}
For $i=k$, using the results in \eqref{E1} to derive
\begin{align}\label{E_DS}
  \mathbb{E}\left\{ {{{\left| {{\text{D}}{{\text{S}}_k}} \right|}^2}} \right\} &= \rho _k^2\left[ {n - \lambda } \right]p_\text{u}{\eta_k}{\left| {\sum\limits_{l = 1}^L {a_{kl}^ * \left[ n \right]\mathbb{E}\left\{ {{\mathbf{\hat h}}_{kl}^{\text{H}}\left[ \lambda  \right] {{{\mathbf{ h}}}_{kl}}\left[ \lambda  \right] } \right\}} } \right|^2} \notag \\
&= \rho _k^2\left[ {n - \lambda } \right]p_\text{u}{\eta_k}{\left| {\sum\limits_{l = 1}^L {a_{kl}^ * \left[ n \right]{\text{tr}}\left( {{{\mathbf{Q}}_{kl}}} \right)} } \right|^2}.
\end{align}

\emph{4) Compute ${\mathbb{E}\left\{ {{{\left| {{\mathrm{B}}{{\mathrm{U}}_{k,n}}} \right|}^2}} \right\}}$:}
The variance of a sum of independent random variables is equal to the sum of the variances. For $i=k$, using the results in \eqref{E1} and \eqref{E2} to obtain
\begin{align}\label{BU1}
  \mathbb{E}\left\{ {{{\left| {{\text{B}}{{\text{U}}_{k,n}}} \right|}^2}} \right\} &= \rho _k^2\left[ {n - \lambda } \right]p_\text{u}{\eta_k}\mathbb{E}\left\{ {{{\left| {\left( {\sum\limits_{l = 1}^L {a_{kl}^ * \left[ n \right]\left( {{\mathbf{\hat h}}_{kl}^{\text{H}}\left[ \lambda  \right]{{\mathbf{h}}_{kl}}\left[ \lambda  \right] - \mathbb{E}\left\{ {{\mathbf{\hat h}}_{kl}^{\text{H}}\left[ \lambda  \right]{{\mathbf{h}}_{kl}}\left[ \lambda  \right]} \right\}} \right)} } \right)} \right|}^2}} \right\} \notag \\
   &= \rho _k^2\left[ {n - \lambda } \right]p_\text{u}{\eta_k}\sum\limits_{l = 1}^L {{{\left| {a_{kl}^ * \left[ n \right]} \right|}^2}{\text{tr}}\left( {{{\mathbf{Q}}_{kl}}{{\mathbf{R}}_{kl}}} \right)} .
\end{align}

\emph{5) Compute ${\mathbb{E}\left\{ {{{\left| {{\mathrm{C}}{{\mathrm{A}}_{k,n}}} \right|}^2}} \right\}}$:}
Using \cite[Eq. (28)]{zheng2020efficient}, we have
\begin{align}\label{q1}
  \mathbb{E}\left\{ {{{\left| {{\text{C}}{{\text{A}}_{k,n}}} \right|}^2}} \right\} &= \bar \rho _k^2\left[ {n - \lambda } \right]p_\text{u}{\eta_k}\mathbb{E}\left\{ {{{\left| {\sum\limits_{l = 1}^L {a_{kl}^ * \left[ n \right]{\mathbf{\hat h}}_{kl}^{\text{H}}\left[ \lambda  \right]{{\mathbf{u}}_{kl}}\left[ n \right]} } \right|}^2}} \right\} \notag \\
  &= \bar \rho _k^2\left[ {n - \lambda } \right]p_\text{u}{\eta_k}\left( {\sum\limits_{l = 1}^L {{{\left| {a_{kl}^ * \left[ n \right]} \right|}^2}{{\Upsilon _3}}} } \right. + \left. {\sum\limits_{l = 1}^L {\sum\limits_{m \ne l}^L {{a_{kl}}\left[ n \right]a_{km}^ * \left[ n \right]{{\Upsilon _4}}} } } \right).
\end{align}
From the definition of channel aging in \eqref{h_il}, ${{\mathbf{u}}_{kl}}\left[ n \right]$ is uncorrelated with ${\mathbf{\hat h}}_{kl}\left[ \lambda  \right]$. We have
\begin{align}
{{\Upsilon _3}} &= \mathbb{E}\left\{ {{{\left| {{\mathbf{\hat h}}_{kl}^{\text{H}}\left[ \lambda  \right]{{\mathbf{u}}_{kl}}\left[ n \right]} \right|}^2}} \right\}={\text{tr}}\left( {{{\mathbf{Q}}_{kl}}{{\mathbf{R}}_{kl}}} \right), \\
{{\Upsilon _4}} &= \mathbb{E}\left\{ {\left( {{\mathbf{\hat h}}_{kl}^{\text{H}}\left[ \lambda  \right]{{\mathbf{u}}_{kl}}\left[ n \right]} \right)\left( {{\mathbf{\hat h}}_{km}^{\text{H}}\left[ \lambda  \right]{{\mathbf{u}}_{km}}\left[ n \right]} \right)} \right\}=0.
\end{align}
Then, substituting above expressions into \eqref{q1}, we obtain
\begin{align}
    \mathbb{E}\left\{ {{{\left| {{\text{C}}{{\text{A}}_{k,n}}} \right|}^2}} \right\}  = \bar \rho _k^2\left[ {n - \lambda } \right]p_\text{u}{\eta_k}\sum\limits_{l = 1}^L {{{\left| {a_{kl}^ * \left[ n \right]} \right|}^2}} {\text{tr}}\left( {{{\mathbf{Q}}_{kl}}{{\mathbf{R}}_{kl}}} \right).
\end{align}

\emph{6) Compute ${\mathbb{E}\left\{ {{{\left| {{\mathrm{U}}{{\mathrm{I}}_{ki,n}}} \right|}^2}} \right\}}$:}
Using \cite[Eq. (28)]{zheng2020efficient}, we have
\begin{align}\label{t}
  \mathbb{E}\left\{ {{{\left| {{\text{U}}{{\text{I}}_{ki,n}}} \right|}^2}} \right\} &= p_\text{u}{\eta_i}\mathbb{E}\left\{ {{{\left| {\sum\limits_{l = 1}^L {a_{kl}^ * \left[ n \right]{\mathbf{\hat h}}_{kl}^{\text{H}}\left[ \lambda  \right]{{\mathbf{h}}_{il}}\left[ n \right]} } \right|}^2}} \right\} \notag \\
   &= p_\text{u}{\eta_i}\sum\limits_{l = 1}^L {{{\left| {a_{kl}^ * \left[ n \right]} \right|}^2}{{\Upsilon _5}} }  + p_\text{u}{\eta_i}\sum\limits_{l = 1}^L {\sum\limits_{m \ne l}^L {{a_{kl}}\left[ n \right]a_{km}^ * \left[ n \right]{{\Upsilon _6}}} }.
\end{align}
By utilizing \eqref{h_il}, we obtain
\begin{align}
\label{t1}{{\Upsilon _5}} &\!=\! \mathbb{E}\!\left\{ {{{\left| {{\mathbf{\hat h}}_{kl}^{\text{H}}\left[ \lambda  \right]\!{{\mathbf{h}}_{il}}\left[ n \right]} \right|}^2}} \right\} \!=\! \rho _i^2\left[ {n \!-\! \lambda } \right]\!\mathbb{E}\!\left\{ {{{\left| {{\mathbf{\hat h}}_{kl}^{\text{H}}\left[ \lambda  \right]\!{{\mathbf{h}}_{il}}\left[ \lambda  \right]} \right|}^2}} \right\} \!+\! \bar \rho _i^2\left[ {n \!-\! \lambda } \right]\!\mathbb{E}\!\left\{ {{{\left| {{\mathbf{\hat h}}_{kl}^{\text{H}}\left[ \lambda  \right]\!{{\mathbf{u}}_{il}}\left[ n \right]} \right|}^2}} \right\},\\
\label{t2}{{\Upsilon _6}} &\!=\! \mathbb{E}\!\left\{\! {{{\left( {{\mathbf{\hat h}}_{kl}^{\text{H}}\!\left[ \lambda  \right]{{\mathbf{h}}_{il}}\!\left[ n \right]} \right)}^*}\!\left( {{\mathbf{\hat h}}_{km}^{\text{H}}\!\left[ \lambda  \right]{{\mathbf{h}}_{im}}\!\left[ n \right]} \right)} \!\right\} \!=\! \rho _i^2\!\left[ {n \!-\! \lambda } \right]\!\mathbb{E}\!\left\{ {{\mathbf{\hat h}}_{kl}^{\text{H}}\!\left[ \lambda  \right]{{\mathbf{h}}_{il}}\!\left[ \lambda  \right]} \right\}\!\mathbb{E}\!\left\{ {{\mathbf{\hat h}}_{km}^{\text{H}}\!\left[ \lambda  \right]{{\mathbf{h}}_{im}}\!\left[ \lambda  \right]} \right\}.
\end{align}
Finally, plugging \eqref{E1}, \eqref{E2}, \eqref{t1} and \eqref{t2} into \eqref{t}, we obtain
\begin{align}\label{E_UI}
  \mathbb{E}\left\{ {{{\left| {{\text{U}}{{\text{I}}_{ki,n}}} \right|}^2}} \right\} \!=\! p_\text{u}{\eta_i}\!\sum\limits_{l = 1}^L {{{\left| {a_{kl}^ * \!\left[ n \right]} \right|}^2}} {\text{tr}}\left( {{{\mathbf{Q}}_{kl}}{{\mathbf{R}}_{il}}} \right) \!+\! \left\{ {\begin{array}{*{20}{c}}
  {\!\rho _i^2\!\left[ {n \!-\! \lambda } \right]p_\text{u}{\eta_i}{{\left| {\sum\limits_{l = 1}^L {a_{kl}^ * \!\left[ n \right]\! {{\text{tr}}\left( {{{{\mathbf{\bar Q}}}_{kil}}} \right)} } } \right|}^2},i \in {\mathcal{P}_k}} \\
  {0,i \notin {\mathcal{P}_k}}
\end{array}} \right.
\end{align}
and this finishes the proof.

\section{Proof of Theorem 2}
When $N=1$, AP $l$ decodes the signal from UE $k$ using only its local estimate ${{\hat h}_{kl}}\left[ {\lambda} \right]$.  The received uplink signal at the $l$th AP is
\begin{align}\label{small cell}
  {y_{{l}}}\left[ n \right] &= {\rho _k}\left[ {n - \lambda } \right]{{\hat h}_{kl}}\left[ \lambda  \right]{s_k}\left[ n \right] + {w_l}\left[ n \right] \notag\\
  &+ \underbrace {{\rho _k}\left[ {n - \lambda } \right]{{\tilde h}_{kl}}\left[ \lambda  \right]{s_k}\left[ n \right] + {{\bar \rho }_k}\left[ {n - \lambda } \right]{\vartheta _{kl}}\left[ n \right]{s_k}\left[ n \right] + \sum\limits_{i \ne k}^K {{h_{il}}\left[ n \right]} {s_i}\left[ n \right]}_{\partial \left[ n \right]},
\end{align}
where ${s_{i}}\left[ {n } \right] \sim \mathcal{C}\mathcal{N}\left( {0,p_\text{u}{\eta _{i}}} \right)$ is transmit power and ${\vartheta_{il}}\left[ {n } \right] \sim \mathcal{C}\mathcal{N}\left( {0,{\beta _{il}}} \right)$ denotes the independent innovation component.
For all ${i \notin {\mathcal{P}_k}}$, ${{{\hat h}_{il}}\left[ {\lambda} \right]}$ and ${{{\hat h}_{kl}}\left[ {\lambda} \right]}$ are independent. For all ${i \in {\mathcal{P}_k}}$, we have
\begin{align}
{{\hat h}_{il}}\left[ {\lambda} \right] = \frac{{{\rho _i}\left[ {\lambda - {t_i}} \right]\sqrt {{p_i}} {\beta _{il}}}}{{{\rho _k}\left[ {\lambda - {t_k}} \right]\sqrt {{p_k}} {\beta _{kl}}}}{{\hat h}_{kl}}\left[ {\lambda} \right].
\end{align}
Using these results, we obtain
\begin{align}\label{partial}
\mathbb{E}\left\{ {{{\left| {\partial \left[ n \right]} \right|}^2}\left| {{{\hat h}_{kl}}\left[ {\lambda} \right]} \right.} \right\} &= \sum\limits_{i \in {\mathcal{P}_k}\setminus \{ k \}}^K {\frac{{\rho _i^2\left[ {n - \lambda} \right]\rho _i^2\left[ {\lambda - {t_i}} \right]p_\text{u}\eta_i p_i\beta _{il}^2}}{{\rho _k^2\left[ {\lambda - {t_k}} \right]{p_k}\beta _{kl}^2}}} \notag\\
&\times {\left| {{{\hat h}_{kl}}\left[ {\lambda} \right]} \right|^2} + p_\text{u}\sum\limits_{i = 1}^K {{\eta_i}{\beta _{il}}}  - p_\text{u}\sum\limits_{i \in {\mathcal{P}_k}}^K {\rho _i^2\left[ {n -\lambda} \right]{\eta_i}{\gamma _{il}}}.
\end{align}
Using the capacity lower bound in \cite[Cor. 1.3]{bjornson2017massive}, an achievable SE at time instant $n$ is
\begin{align}
  &\mathbb{E}\left\{ {{{\log }_2}\left( {1 + {{\left| {{{\hat h}_{kl}}\left[ {\lambda} \right]} \right|}^2}\frac{{\rho _k^2\left[ {n - \lambda} \right]p_\text{u}{\eta_k}\left( {1 + {A_{kl}}\left[ n \right]} \right)}}{{p_\text{u}\sum\limits_{i = 1}^K {{\eta_i}{\beta _{il}}}  -  p_\text{u}\sum\limits_{i \in {\mathcal{P}_k}}^K {\rho _i^2\left[ {n - \lambda} \right]{\eta_i}{\gamma _{il}}}  + {\sigma^2}}}} \right)} \right\} \notag \\
   &- \mathbb{E}\left\{ {{{\log }_2}\left( {1 + {{\left| {{{\hat h}_{kl}}\left[ {\lambda} \right]} \right|}^2}\frac{{\rho _k^2\left[ {n - \lambda} \right]p_\text{u}{\eta_k}{A_{kl}}\left[ n \right]}}{{p_\text{u}\sum\limits_{i = 1}^K {{\eta_i}{\beta _{il}}}  - p_\text{u}\sum\limits_{i \in {\mathcal{P}_k}}^K {\rho _i^2\left[ {n - \lambda} \right]{\eta_i}{\gamma _{il}}}  + {\sigma^2}}}} \right)} \right\}. \notag
\end{align}
Then, with the help of \cite[Lemma 3]{bjornson2010cooperative}, we can compute each of the expectations to obtain the final expression in \eqref{w}.

\section{Proof of Theorem 3}
Here, we use successive interference cancellation technology. At the beginning of the detection process, UE $k$ does not know any of the transmitted signals. It first detects the signal from AP $l$ by using the average channel $\mathbb{E}\left\{ {{\mathbf{h}}_{k1}^{\text{H}}\left[ \lambda  \right]{{{\mathbf{\hat h}}}_{k1}}\left[ \lambda  \right]} \right\}$ only. The received signal can be written as
\begin{align}
  r_{k,1}^{{\text{nc}}}\left[ n \right] &= r_k^{{\text{nc}}}\left[ n \right] = {\rho _k}\left[ {n - \lambda } \right]\sqrt {{p_{\text{d}}}} \mathbb{E}\left\{ {{\mathbf{h}}_{k1}^{\text{H}}\left[ \lambda  \right]{{{\mathbf{\hat h}}}_{k1}}\left[ \lambda  \right]} \right\}\sqrt{\mu _{k1}}{q_{k1}}\left[ n \right] \notag \\
   &+ {\rho _k}\left[ {n - \lambda } \right]\sqrt {{p_{\text{d}}}} \left( {{\mathbf{h}}_{k1}^{\text{H}}\left[ \lambda  \right]{{{\mathbf{\hat h}}}_{k1}}\left[ \lambda  \right] - \mathbb{E}\left\{ {{\mathbf{h}}_{k1}^{\text{H}}\left[ \lambda  \right]{{{\mathbf{\hat h}}}_{k1}}\left[ \lambda  \right]} \right\}} \right)\sqrt{\mu _{k1}}{q_{k1}}\left[ n \right] \notag \\
   &+ {{\bar \rho }_k}\left[ {n - \lambda } \right]\sqrt {{p_{\text{d}}}} {\mathbf{u}}_{k1}^{\text{H}}\left[ n \right]{{{\mathbf{\hat h}}}_{k1}}\left[ \lambda  \right]\sqrt{\mu _{k1}}{q_{k1}}\left[ n \right] + \sqrt{p_{\text{d}}}\sum\limits_{j = 2}^L {{\mathbf{h}}_{kj}^{\text{H}}\left[ n \right]{{{\mathbf{\hat h}}}_{kj}}\left[ \lambda  \right]\sqrt {{\mu _{kj}}} {q_{kj}}\left[ n \right]}  \notag \\
   &+ \sqrt{p_{\text{d}}}\sum\limits_{i \ne k}^K {\sum\limits_{j = 1}^L {{\mathbf{h}}_{kj}^{\text{H}}\left[ n \right]{{{\mathbf{\hat h}}}_{ij}}\left[ \lambda  \right]\sqrt {{\mu _{ij}}} {q_{ij}}\left[ n \right]} }  + {w_k}\left[ n \right] ,
\end{align}
where the first term is the desired signal over known deterministic channel while other terms are treated as uncorrelated noise.
Sequentially, UE $k$ detects signal from AP $l$ by subtracting the first $l-1$ signals:
\begin{align}\label{r_km}
  r_{k,l}^{{\text{nc}}}\left[ n \right] &= r_k^{{\text{nc}}}\left[ n \right] - {\rho _k}\left[ {n - \lambda } \right]\sqrt{p_{\text{d}}}\sum\limits_{j = 1}^{l - 1} { \mathbb{E}\left\{ {{\mathbf{h}}_{kj}^{\text{H}}\left[ \lambda  \right]{{{\mathbf{\hat h}}}_{kj}}\left[ \lambda  \right]} \right\}\sqrt {{\mu _{kj}}}{q_{kj}}\left[ n \right]}  \notag \\
   &= {\rho _k}\left[ {n - \lambda } \right]\sqrt {{p_{\text{d}}}} \mathbb{E}\left\{ {{\mathbf{h}}_{kl}^{\text{H}}\left[ \lambda  \right]{{{\mathbf{\hat h}}}_{kl}}\left[ \lambda  \right]} \right\}\sqrt{\mu _{kl}}{q_{kl}}\left[ n \right] \notag \\
   &+ {\rho _k}\left[ {n - \lambda } \right]\sqrt{p_{\text{d}}}\sum\limits_{j = 1}^l {\left( {{\mathbf{h}}_{kj}^{\text{H}}\left[ \lambda  \right]{{{\mathbf{\hat h}}}_{kj}}\left[ \lambda  \right] - \mathbb{E}\left\{ {{\mathbf{h}}_{kj}^{\text{H}}\left[ \lambda  \right]{{{\mathbf{\hat h}}}_{kj}}\left[ \lambda  \right]} \right\}} \right)\sqrt {{\mu _{kj}}} {q_{kj}}\left[ n \right]}  \notag \\
   &+ {{\bar \rho }_k}\left[ {n - \lambda } \right]\sqrt{p_{\text{d}}}\sum\limits_{j = 1}^l { {\mathbf{u}}_{kj}^{\text{H}}\left[ n \right]{{{\mathbf{\hat h}}}_{kj}}\left[ \lambda  \right]\sqrt {{\mu _{kj}}}{q_{kj}}\left[ n \right]}  + \sqrt{p_{\text{d}}} \sum\limits_{j = l + 1}^L {{\mathbf{h}}_{kj}^{\text{H}}\left[ n \right]{{{\mathbf{\hat h}}}_{kj}}\left[ \lambda  \right]\sqrt {{\mu _{kj}}} {q_{kj}}\left[ n \right]}  \notag \\
   &+ \sqrt{p_{\text{d}}}\sum\limits_{i \ne k}^K {\sum\limits_{j = 1}^L {{\mathbf{h}}_{kj}^{\text{H}}\left[ n \right]{{{\mathbf{\hat h}}}_{ij}}\left[ \lambda  \right]\sqrt {{\mu _{ij}}} {q_{ij}}\left[ n \right]} }  + {w_k}\left[ n \right] .
\end{align}
The first term in \eqref{r_km} is equivalent to having a deterministic channel
\begin{align}
h\left[ n \right]={\rho _k}\left[ {n - \lambda } \right]\sqrt {{p_{\text{d}}}{\mu _{kl}}} \mathbb{E}\left\{ {{\mathbf{h}}_{kl}^{\text{H}}\left[ \lambda  \right]{{{\mathbf{\hat h}}}_{kl}}\left[ \lambda  \right]} \right\},
\end{align}
and ${q_{kl}}\left[ n \right]$ is the desired signal. The other terms are uncorrelated noise $v_{kl}\left[ n \right]$ with power
\begin{align}
  \mathbb{E}\left\{ {{{\left| {{v_{kl}\left[ n \right]}} \right|}^2}} \right\}
   &= {p_{\text{d}}}\sum\limits_{i = 1}^K {\sum\limits_{j = 1}^L {{\mu _{ij}}\mathbb{E}\left\{ {{{\left| {{\mathbf{h}}_{kj}^{\text{H}}\left[ n \right]{{{\mathbf{\hat h}}}_{ij}}\left[ \lambda  \right]} \right|}^2}} \right\}} }  \notag\\
   &- \rho _k^2\left[ {n - \lambda } \right]{p_{\text{d}}}\sum\limits_{j = 1}^l {{\mu _{kj}}{{\left| {\mathbb{E}\left\{ {{\mathbf{h}}_{kj}^{\text{H}}\left[ \lambda  \right]{{{\mathbf{\hat h}}}_{kj}}\left[ \lambda  \right]} \right\}} \right|}^2}}  + \sigma _{\text{d}}^2 .
\end{align}
The downlink SINR at time instant $n$ is equal to ${\zeta _{kl}\left[ n \right]} = {{{{\left| h\left[ n \right] \right|}^2}}}/{{\mathbb{E}\left\{ {{{\left| {{v_{kl}\left[ n \right]}} \right|}^2}} \right\}}}$ as
\begin{align}
  \frac{{\rho _k^2\left[ {n - \lambda } \right]{p_{\text{d}}}{\mu _{kl}}{{\left| {\mathbb{E}\left\{ {{\mathbf{h}}_{kl}^{\text{H}}\left[ \lambda  \right]{{{\mathbf{\hat h}}}_{kl}}\left[ \lambda  \right]} \right\}} \right|}^2}}}{{{p_{\text{d}}}\sum\limits_{i = 1}^K {\sum\limits_{j = 1}^L {{\mu _{ij}}\mathbb{E}\left\{ {{{\left| {{\mathbf{h}}_{kj}^{\text{H}}\left[ n \right]{{{\mathbf{\hat h}}}_{ij}}\left[ \lambda  \right]} \right|}^2}} \right\}} }  - \rho _k^2\left[ {n - \lambda } \right]{p_{\text{d}}}\sum\limits_{j = 1}^l {{\mu _{kj}}{{\left| {\mathbb{E}\left\{ {{\mathbf{h}}_{kj}^{\text{H}}\left[ \lambda  \right]{{{\mathbf{\hat h}}}_{kj}}\left[ \lambda  \right]} \right\}} \right|}^2}}  + \sigma _{\text{d}}^2}}.
\end{align}
Then, the total SE of UE $k$ at time instant $n$ is equal to
\begin{align}\label{SE_k}
  &{\text{S}}{{\text{E}}_k}\left[ n \right] = {\log _2}\left( {\prod\limits_{l = 1}^L {\left( {1 + {\zeta _{kl}}\left[ n \right]} \right)} } \right) \notag \\
   &= {\log _2}\!\left(\! {\frac{{{p_{\text{d}}}\sum\limits_{i = 1}^K {\sum\limits_{j = 1}^L {{\mu _{ij}}\mathbb{E}\left\{ {{{\left| {{\mathbf{h}}_{kj}^{\text{H}}\left[ n \right]{{{\mathbf{\hat h}}}_{ij}}\left[ \lambda  \right]} \right|}^2}} \right\}} }  + \sigma _{\text{d}}^2}}{{{p_{\text{d}}}\!\sum\limits_{i = 1}^K {\!\sum\limits_{j = 1}^L {{\mu _{ij}}\mathbb{E}\!\left\{ {{{\left| {{\mathbf{h}}_{kj}^{\text{H}}\!\left[ n \right]{{{\mathbf{\hat h}}}_{ij}}\!\left[ \lambda  \right]} \right|}^2}} \right\}} }  \!-\! \rho _k^2\!\left[ {n \!-\! \lambda } \right]{p_{\text{d}}}\!\sum\limits_{j = 1}^L {{\mu _{kj}}{{\left| {\mathbb{E}\!\left\{ {{\mathbf{h}}_{kj}^{\text{H}}\!\left[ \lambda  \right]{{{\mathbf{\hat h}}}_{kj}}\!\left[ \lambda  \right]} \right\}} \right|}^2}}  \!+\! \sigma _{\text{d}}^2}}} \!\right)\!.
\end{align}
According to \eqref{SE_k}, so we can obtain \eqref{SE_c}.
\end{appendices}

\bibliographystyle{IEEEtran}
\bibliography{IEEEabrv,Ref}

\begin{thebibliography}{10}
\providecommand{\url}[1]{#1}
\csname url@samestyle\endcsname
\providecommand{\newblock}{\relax}
\providecommand{\bibinfo}[2]{#2}
\providecommand{\BIBentrySTDinterwordspacing}{\spaceskip=0pt\relax}
\providecommand{\BIBentryALTinterwordstretchfactor}{4}
\providecommand{\BIBentryALTinterwordspacing}{\spaceskip=\fontdimen2\font plus
\BIBentryALTinterwordstretchfactor\fontdimen3\font minus
  \fontdimen4\font\relax}
\providecommand{\BIBforeignlanguage}[2]{{%
\expandafter\ifx\csname l@#1\endcsname\relax
\typeout{** WARNING: IEEEtran.bst: No hyphenation pattern has been}%
\typeout{** loaded for the language `#1'. Using the pattern for}%
\typeout{** the default language instead.}%
\else
\language=\csname l@#1\endcsname
\fi
#2}}
\providecommand{\BIBdecl}{\relax}
\BIBdecl

\bibitem{zhang2020prospective}
J.~Zhang, E.~Bj{\"o}rnson, M.~Matthaiou, D.~W.~K. Ng, H.~Yang, and D.~J. Love,
  ``Prospective multiple antenna technologies for beyond 5g,'' \emph{IEEE J.
  Sel. Areas Commun.}, vol.~38, no.~8, pp. 1637--1660, Aug. 2020.

\bibitem{Ngo2017Cell}
H.~Q. Ngo, A.~Ashikhmin, Y.~Hong, E.~G. Larsson, and T.~L. Marzetta,
  ``Cell-free massive mimo versus small cells,'' \emph{IEEE Trans. Wireless
  Commun.}, vol.~16, no.~3, pp. 1834--1850, Mar. 2017.

\bibitem{8000355}
S.~{Buzzi} and C.~{D'Andrea}, ``Cell-free massive {MIMO}: {U}ser-centric
  approach,'' \emph{IEEE Wireless Commun. Lett.}, vol.~6, no.~6, pp. 706--709,
  2017.

\bibitem{9186090}
X.~{Zhang}, J.~{Wang}, and H.~V. {Poor}, ``Statistical delay and error-rate
  bounded {QoS} provisioning for {mURLLC} over {6G CF M-MIMO} mobile networks
  in the finite blocklength regime,'' \emph{IEEE J. Sel. Areas Commun.}, to
  appear, 2020.

\bibitem{9184916}
F.~{Tan}, P.~{Wu}, Y.~{Wu}, and M.~{Xia}, ``Energy-efficient non-orthogonal
  multicast and unicast transmission of cell-free massive {MIMO} systems with
  {SWIPT},'' \emph{IEEE J. Sel. Areas Commun.}, to appear, 2020.

\bibitem{bjornson2019making}
E.~Bj{\"o}rnson and L.~Sanguinetti, ``Making cell-free massive mimo competitive
  with mmse processing and centralized implementation,'' \emph{IEEE Trans.
  Wireless Commun.}, vol.~19, no.~1, pp. 77--90, Jan. 2020.

\bibitem{nayebi2017precoding}
E.~Nayebi, A.~Ashikhmin, T.~L. Marzetta, H.~Yang, and B.~D. Rao, ``{Precoding
  and power optimization in cell-free massive MIMO systems},'' \emph{IEEE
  Trans. Wireless Commun.}, vol.~16, no.~7, pp. 4445--4459, Jul. 2017.

\bibitem{zheng2020efficient}
J.~Zheng, J.~Zhang, L.~Zhang, X.~Zhang, and B.~Ai, ``Efficient receiver design
  for uplink cell-free massive mimo with hardware impairments,'' \emph{IEEE
  Trans. Veh. Technol.}, vol.~69, no.~4, pp. 4537--4541, Apr. 2020.

\bibitem{qiu2020downlink}
J.~Qiu, K.~Xu, X.~Xia, Z.~Shen, and W.~Xie, ``{Downlink power optimization for
  cell-free massive MIMO over spatially correlated Rayleigh fading channels},''
  \emph{IEEE Access}, vol.~8, pp. 56\,214--56\,227, 2020.

\bibitem{interdonato2019ubiquitous}
G.~Interdonato, E.~Bj{\"o}rnson, H.~Q. Ngo, P.~Frenger, and E.~G. Larsson,
  ``{Ubiquitous cell-free massive MIMO communications},'' \emph{EURASIP J.
  Wireless Commun. Netw.}, vol. 2019, no.~1, p. 197, Jan. 2019.

\bibitem{9079911}
T.~C. {Mai}, H.~Q. {Ngo}, and T.~Q. {Duong}, ``{Downlink spectral efficiency of
  cell-free massive MIMO systems with multi-antenna users},'' \emph{IEEE Trans.
  Commun.}, vol.~68, no.~8, pp. 4803--4815, Aug. 2020.

\bibitem{9178782}
S.~{Buzzi}, C.~{D¡¯Andrea}, M.~{Fresia}, Y.~P. {Zhang}, and S.~{Feng}, ``{Pilot
  assignment in cell-free massive MIMO based on the Hungarian algorithm},''
  \emph{IEEE Wireless Commun. Lett.}, vol.~10, no.~1, pp. 34--37, Jan. 2021.

\bibitem{mai2018pilot}
T.~C. Mai, H.~Q. Ngo, M.~Egan, and T.~Q. Duong, ``{Pilot power control for
  cell-free massive MIMO},'' \emph{IEEE Trans. Veh. Technol.}, vol.~67, no.~11,
  pp. 11\,264--11\,268, Nov. 2018.

\bibitem{fozooni2019hybrid}
M.~Fozooni, H.~Q. Ngo, M.~Matthaiou, S.~Jin, and G.~C. Alexandropoulos,
  ``{Hybrid processing design for multipair massive MIMO relaying with channel
  spatial correlation},'' \emph{IEEE Trans. Commun.}, vol.~67, no.~1, pp.
  107--123, Jan. 2019.

\bibitem{polegre2020channel}
A.~{\'A}. Polegre, F.~Riera-Palou, G.~Femenias, and A.~G. Armada, ``{Channel
  hardening in cell-free and user-centric massive MIMO networks with spatially
  correlated Ricean fading},'' \emph{IEEE Access}, vol.~8, pp.
  139\,827--139\,845, 2020.

\bibitem{jin2020spectral}
S.-N. Jin, D.-W. Yue, and H.~H. Nguyen, ``{Spectral and energy efficiency in
  cell-free massive MIMO systems over correlated Rician fading},'' \emph{IEEE
  Systems Journal}, to appear, 2020.

\bibitem{bjornson2017massive}
E.~Bj{\"o}rnson, J.~Hoydis, and L.~Sanguinetti, ``{Massive MIMO networks:
  Spectral, energy, and hardware efficiency},'' \emph{Foundations and
  Trends{\textregistered} in Signal Processing}, vol.~11, no. 3-4, pp.
  154--655, 2017.

\bibitem{9145564}
I.~F. {Akyildiz}, A.~{Kak}, and S.~{Nie}, ``{6G} and beyond: {T}he future of
  wireless communications systems,'' \emph{IEEE Access}, vol.~8, pp.
  133\,995--134\,030, 2020.

\bibitem{9103348}
B.~{Ai}, A.~F. {Molisch}, M.~{Rupp}, and Z.~{Zhong}, ``{5G} key technologies
  for smart railways,'' \emph{Proc. IEEE}, vol. 108, no.~6, pp. 856--893, May
  2020.

\bibitem{bjornson2015optimal}
E.~Bj{\"o}rnson, L.~Sanguinetti, J.~Hoydis, and M.~Debbah, ``{Optimal design of
  energy-efficient multi-user MIMO systems: Is massive MIMO the answer?}''
  \emph{IEEE Trans. Wireless Commun.}, vol.~14, no.~6, pp. 3059--3075, Jun.
  2015.

\bibitem{9124715}
T.~T. {Vu}, D.~T. {Ngo}, N.~H. {Tran}, H.~Q. {Ngo}, M.~N. {Dao}, and R.~H.
  {Middleton}, ``Cell-free massive {MIMO} for wireless federated learning,''
  \emph{IEEE Trans. Wireless Commun.}, to appear, 2020.

\bibitem{ngo2017total}
H.~Q. Ngo, L.-N. Tran, T.~Q. Duong, M.~Matthaiou, and E.~G. Larsson, ``{On the
  total energy efficiency of cell-free massive MIMO},'' \emph{IEEE Trans. Green
  Commun. Netw.}, vol.~2, no.~1, pp. 25--39, Jan. 2017.

\bibitem{bashar2019energy}
M.~Bashar, K.~Cumanan, A.~G. Burr, H.~Q. Ngo, E.~G. Larsson, and P.~Xiao,
  ``Energy efficiency of the cell-free massive mimo uplink with optimal uniform
  quantization,'' \emph{IEEE Trans. Green Commun. Netw.}, vol.~3, no.~4, pp.
  971--987, Apr. 2019.

\bibitem{van2020joint}
T.~Van~Chien, E.~Bj{\"o}rnson, and E.~G. Larsson, ``{Joint power allocation and
  load balancing optimization for energy-efficient cell-free massive MIMO
  networks},'' \emph{IEEE Trans. Wireless Commun.}, to appear, 2020.

\bibitem{truong2013effects}
K.~T. Truong and R.~W. Heath, ``{Effects of channel aging in massive MIMO
  systems},'' \emph{J. Commun. Netw.}, vol.~15, no.~4, pp. 338--351, Apr. 2013.

\bibitem{yuan2020machine}
J.~Yuan, H.~Q. Ngo, and M.~Matthaiou, ``{Machine learning-based channel
  prediction in massive MIMO with channel aging},'' \emph{IEEE Trans. Wireless
  Commun.}, vol.~19, no.~5, pp. 2960--2973, May 2020.

\bibitem{chopra2018performance}
R.~Chopra, C.~R. Murthy, H.~A. Suraweera, and E.~G. Larsson, ``{Performance
  analysis of FDD massive MIMO systems under channel aging},'' \emph{IEEE
  Trans. Wireless Commun.}, vol.~17, no.~2, pp. 1094--1108, Feb. 2018.

\bibitem{papazafeiropoulos2016impact}
A.~K. Papazafeiropoulos, ``{Impact of general channel aging conditions on the
  downlink performance of massive MIMO},'' \emph{IEEE Trans. Veh. Technol.},
  vol.~66, no.~2, pp. 1428--1442, Feb. 2016.

\bibitem{8716688}
S.~{Wu}, E.~{Bj{\"o}rnson}, C.~{Moll¨¦n}, X.~{Tao}, and E.~G. {Larsson},
  ``{Inverse extrapolation for efficient precoding in time-varying massive
  MIMO-OFDM systems},'' \emph{IEEE Access}, vol.~7, pp. 91\,105--91\,119, 2019.

\bibitem{zheng2020cell}
J.~Zheng, J.~Zhang, E.~Bjornson, and B.~Ai, ``{Cell-free massive MIMO with
  channel aging and pilot contamination},'' in \emph{Proc. IEEE GLOBECOM},
  2020, pp. 1--6.

\bibitem{Abramowitz1964table}
M.~Abramowitz and I.~A. Stegun, \emph{{Handbook of Mathematical Functions With
  Formulas, Graphs, and Mathematical Tables}}, 9th~ed.\hskip 1em plus 0.5em
  minus 0.4em\relax New York, NY, USA: Dover, 1964.

\bibitem{nikbakht2019uplink}
R.~Nikbakht and A.~Lozano, ``{Uplink fractional power control for cell-free
  wireless networks},'' in \emph{Proc. IEEE ICC}, May 2019, pp. 1--5.

\bibitem{interdonato2019scalability}
G.~Interdonato, P.~Frenger, and E.~G. Larsson, ``{Scalability aspects of
  cell-free massive MIMO},'' in \emph{Proc. IEEE ICC}, May 2019, pp. 1--6.

\bibitem{kashyap2016frequency}
S.~Kashyap, C.~Moll{\'e}n, E.~Bj{\"o}rnson, and E.~G. Larsson,
  ``{Frequency-domain interpolation of the zero-forcing matrix in massive
  MIMO-OFDM},'' in \emph{Proc. IEEE SPAWC}, 2016, pp. 1--5.

\bibitem{bjornson2010cooperative}
E.~Bj{\"o}rnson, R.~Zakhour, D.~Gesbert, and B.~Ottersten, ``{Cooperative
  multicell precoding: Rate region characterization and distributed strategies
  with instantaneous and statistical CSI},'' \emph{IEEE Trans. Signal
  Process.}, vol.~58, no.~8, pp. 4298--4310, Aug. 2010.

\end{thebibliography}

\end{document}